%% file: paper160713.tex
\documentclass[11pt,epsfig,number,sort&compress]{article}
\usepackage{graphicx,color}
\usepackage{appendix}
\usepackage{latexsym,amsmath,amssymb,graphicx,booktabs}
\usepackage{epsfig,latexsym,cite}
\usepackage{hyperref}
\numberwithin{equation}{section}

\definecolor{MyBlue}{rgb}{0.15,0.15,0.70}

\hypersetup{
colorlinks=true,
citecolor=MyBlue,
linkcolor=MyBlue,
urlcolor=MyBlue
}

\input epsf
\usepackage{amssymb}
\usepackage{amsmath}
\usepackage{amsfonts}
\usepackage{upgreek}
\usepackage{latexsym}

\include{mydefs}

\begin{document}

\begin{titlepage}

\vspace*{2cm}

\centerline{\Large \bf A non-local theory  of  massive gravity
}
\vskip 0.4cm
\vskip 0.7cm
\centerline{\large Maud Jaccard, Michele Maggiore and Ermis Mitsou}
\vskip 0.3cm
\centerline{\em D\'epartement de Physique Th\'eorique and Center for Astroparticle Physics,}  
\centerline{\em Universit\'e de Gen\`eve, 24 quai Ansermet, CH--1211 Gen\`eve 4, Switzerland}

\vskip 1.9cm

\begin{abstract}

We construct a fully covariant theory of massive gravity which does not require the introduction  of an external reference metric, and  overcomes  the usual problems   of massive gravity theories (fatal ghosts instabilities, acausality and/or vDVZ discontinuity). The equations of motion of the theory  are non-local,  but respect causality.
The starting point is the quadratic action proposed in the context
of the degravitation idea. 
We show that it is possible to extended it to a fully non-linear covariant theory.   
This theory describes the five degrees of freedom of a massive graviton plus a scalar ghost. However, contrary to generic non-linear extensions of Fierz-Pauli massive gravity, the ghost has the same mass $m$ as the massive graviton, independently of the  background, and smoothly goes into  a
non-radiative  degree of freedom for $m\ra 0$. As a consequence,
for $m\sim H_0$ the vacuum instability induced by the ghost is irrelevant even over  cosmological time-scales.   
We  finally  show that an extension of the model  degravitates a vacuum energy density of order $\mpl^4$ down to a value of order $\mpl^2 m^2$, which  for $m={\cal O}(H_0)$ is of order of the observed value of the vacuum energy density.

\end{abstract}


\end{titlepage}

\newpage

\section{Introduction and Summary}
The search for a viable theory of massive gravity provides a long-standing challenge to theoretical physics, and has a long history \cite{Fierz:1939ix,Boulware:1973my}.
Recent years have witnessed an explosion of activity on the subject, and more generally on infrared modifications of GR,  motivated both by the intrinsic field-theoretical interest of the problem and by its potential relevance for understanding the origin of dark energy. This has lead to beautiful theoretical ideas such as the DGP model~\cite{Dvali:2000hr}, 
degravitation of vacuum energy~\cite{Dvali:2000xg,Dvali:2002pe,ArkaniHamed:2002fu,Dvali:2006su,Dvali:2007kt}, effective field theories for massive gravity based on the \Stu mechanism \cite{ArkaniHamed:2002sp},
galileons~\cite{Nicolis:2008in},  and  to the construction of a ghost-free theory of massive gravity, the dRGT theory
\cite{deRham:2010ik,deRham:2010kj} (see also \cite{deRham:2010gu,deRham:2011rn,deRham:2011qq,Hassan:2011hr,Hassan:2011vm,Hassan:2011tf,Hassan:2011ea,Hassan:2012qv,Comelli:2012vz,Comelli:2013txa}, and ref.~\cite{Hinterbichler:2011tt} for  a review). Cosmological consequences of these ideas have been extensively explored. 
A self-accelerated solution was first found in the context of DGP~\cite{Deffayet:2000uy,Deffayet:2001pu}, but is unfortunately   plagued by a 
ghost instability~\cite{Luty:2003vm,Nicolis:2004qq,Gorbunov:2005zk,Charmousis:2006pn,Izumi:2006ca}.
Self-accelerated solutions, with the Hubble parameter determined by the graviton mass, have also been found in dRGT theory \cite{deRham:2010tw,Koyama:2011xz,Koyama:2011yg,Nieuwenhuizen:2011sq,Chamseddine:2011bu,D'Amico:2011jj,DeFelice:2013bxa,Tasinato:2013rza}.

Despite these remarkable advances, some crucial problems remain open. In dRGT,
even if the sixth ghost-like degree of freedom is absent in any background, the fluctuations of the remaining five degrees of freedom can become ghost-like over non-trivial backgrounds. In particular,   the self-accelerating solutions of dRGT theory generically have scalar or vector instabilities \cite{Gumrukcuoglu:2011ew,Gumrukcuoglu:2011zh,Koyama:2011wx}.
Another important open problem is posed by the existence of  superluminal 
modes over some backgrounds \cite{Dubovsky:2005xd,Gruzinov:2011sq,deRham:2011pt,Deser:2012qx,Berezhiani:2013dw,Izumi:2013poa,Berezhiani:2013dca,Koyama:2013paa}, which also  appears in galileon theories
\cite{Nicolis:2009qm}. As discussed in \cite{Burrage:2011cr} in the case of galileons, this however does not necessarily implies the loss of causality, since in the attempt of constructing closed time-like curves one is forced to leave the domain of validity of the effective field theory.

At a different (and possibly more subjective) level, the need for an external reference metric in dRGT theory is disturbing. This can be seen particularly clearly in the unitary gauge, where the \Stu fields $\phi^a$ are set to zero. In this gauge the theory is constructed in terms of a field $\hmn=\gmn-\gbmn$, where $\gbmn$ is a fiducial reference metric given by a classical solution of Einstein gravity. In practice, this means that we have a different Lagrangian for the massive theory for any classical solution of the massless theory, a situation that can hardly be accepted as fundamental. Bimetric versions of ghost-free massive gravity 
\cite{Hassan:2011zd,Hassan:2012wr,Hassan:2012gz,Comelli:2012db,Berg:2012kn}
address this concern by assigning a dynamics to the reference metric, but this seems to spoil, or at least significantly complicate, the simple and beautiful geometric interpretation of GR.

In this paper we approach the problem of constructing a consistent theory of massive gravity from a different point of view, developing a non-local formulation of the theory. One might fear that non-locality brings in new conceptual complications.  However,
while it is true that  non-locality can bring in technical complications (e.g. integro-differential equations of motion), conceptual issues that are sometimes raised in this context are actually rather  due to some common misconceptions about non-local theories. For instance, one should not mix up non-locality with lack of causality. If in an equation of motion there is a term proportional to  the inverse of the d'Alembertian operator $\iBox$, this does not mean that the theory is acausal, as long as 
$\iBox$ is defined in terms of the retarded Green's function. Another common misconception is that non-local theories necessarily hide extra ghost-like degrees of freedom, much as higher-derivative theories. If the equations of motion involve a function
$f(\Box)$ and we  expand this function and truncate the expansion to a finite order $N$, we  indeed have a higher-derivative theory with time derivatives up to  order $2N$. This requires $2N$ data as initial conditions and therefore describes $N$ degrees of freedom. The   Ostrogradski theorem ensures that at least one of these extra degrees of freedom is a  ghost. 
However, as discussed  in \cite{Eliezer:1989cr}
(see also \cite{Simon:1990ic,Woodard:2006nt,Barnaby:2007ve}),  in general
the solutions of the truncated theory are spurious and do not converge to solutions of the full non-local theory as the order of the expansion $N\ra\infty$. In particular when $f(\Box)$ is non-analytic, e.g.  $f(\Box)=1/(\Box-m^2)$, most of the solutions of the truncated theory have large frequencies, which lie outside the convergence radius of the derivative expansion, and for $N\ra \infty$ they do not converge to solutions of the full theory.
Non-local theories with non-analytic functions $f(z)$
emerge for instance from the integration of  degrees of freedom in a perfectly healthy theory, so in this case it is  clear that they have  no pathology. As a trivial example, one can consider the non-local
Lagrangian \cite{Eliezer:1989cr}
\be
L[q]=\frac{1}{2}\dot{q}^2-\frac{1}{2}\omega^2q^2+
\frac{1}{2}g^2\omega^2q\(\frac{\omega^2}{d^2/dt^2 +\omega^2}\) q\, .
\ee
An expansion of the non-local factor in powers of $d^2/dt^2$, followed by truncation to a finite order $N$, leads to a theory that requires $2N$ initial data and so describes $N$ degrees of freedom, out of which   at least one is an Ostrogradski ghost. However, the Lagrangian
$L[q]$
can be obtained by  integrating out the $x$ variable from the Lagrangian
\be
L'[q,x]=\frac{1}{2}\dot{q}^2-\frac{1}{2}\omega^2q^2 +
\frac{1}{2}\dot{x}^2-\frac{1}{2}\omega^2x^2 -g\omega^2xq\, ,
\ee
which describes two coupled harmonic oscillators, and obviously has no pathologies. 
Similar examples also  commonly occur in quantum field theory. For instance, even if
the running of coupling constants is more frequently formulated in momentum space, there is also  an alternative formulation, developed  in gauge theories and in quantum gravity in the pioneering works
\cite{Barvinsky:1985an,Barvinsky:1987uw} (see also \cite{Ostrovsky:1988jn,Buchbinder:1992rb,Dobado:1998mr,Barvinsky:2003kg,Shapiro:2008sf,Shapiro:2009dh}), which
uses non-local effective actions in coordinate space. In this  formalism 
the one-loop effective action obtained in  QED  by integrating out the fermionic fields can be written in  the form
\be
S_{\rm eff}=-\frac{1}{4}\int d^4x\, \Fmn\[ \frac{1}{g_0^2}
+\beta_0\ln\(\frac{-\Box}{\mu^2}\)\]\FMN\, ,
\ee
where $\beta_0$ is the one-loop $\beta$-function   and $\mu$ an appropriately chosen mass  scale.\footnote{The operator $\ln(-\Box/\mu^2)$ can be defined via a momentum space convolution or, equivalently, from
$$
\ln\(\frac{-\Box}{\mu^2}\)=\int_0^{\infty}dm^2\, \[
\frac{1}{m^2+\mu^2}-\frac{1}{m^2-\Box}\]\, .
$$}
Non-local expressions  can also emerge from the reduction to four dimensions of higher-dimensional theories, as in DGP, where the reduction to four dimensions gives an action involving $\sqrt{-\Box}$ \cite{Dvali:2000hr,Hinterbichler:2011tt}.
Thus, non-local theories do not necessarily have pathologies, and non-local modifications of gravity have been discussed in a number of different contexts, in the attempt to construct both IR~\cite{ArkaniHamed:2002fu,Dvali:2006su,Dvali:2007kt}
and UV modifications~\cite{Hamber:2005dw,Khoury:2006fg,Modesto:2011kw,Briscese:2012ys}. Non-local operators also naturally enter in the description of fields with spin $s>2$~\cite{Francia:2002aa}.

We will see in this paper that the use of a non-local formulation can be very useful in massive gauge theories and in massive gravity. Indeed, at the level of linearized theories there is a sort of duality between gauge invariance and locality, which can both be made manifest in the formalism, but in a mutually exclusive manner. The simplest example is given by massive electrodynamics, governed by the Proca Lagrangian,
\be\label{calL1}
{\cal L}=-\frac{1}{4}F_{\mu\nu}F^{\mu\nu}-\frac{1}{2}m_{\g}^2A_{\mu}A^{\mu}\, .
\ee
As we will recall in Sect.~\ref{sect:massiveQED} (following refs.~\cite{Dvali:2006su,Dvali:2007kt}), this theory is actually equivalent to a theory with Lagrangian
\be\label{calL2}
{\cal L}'=
-\frac{1}{4} \Fmn \(1-\frac{m_{\g}^2}{\Box}\)\FMN\, .
\ee
The formulation (\ref{calL1}) is explicitly local, but not gauge-invariant. In contrast, the Lagrangian (\ref{calL2}) is explicitly gauge-invariant, even in the massive case, at the price of manifest locality.  Observe that the equation of motion derived from \eq{calL2} can be written as 
\be
(\Box-m_{\g}^2)\AN=\(1-\frac{m_{\g}^2}{\Box}\)\paN\pam\AMU\, .
\ee
The non-local term in this equation  can be eliminated  by fixing the gauge $\pam A^{\mu}=0$, thereby recovering the equations of motion derived directly from (\ref{calL1}). Thus the theory described by (\ref{calL1}),  which is not gauge-invariant, can be understood as the gauge fixing of a gauge-invariant but non-manifestly local theory. Fixing the gauge we lose manifest gauge invariance, but at the same time we eliminate the non-local terms.

The same strategy can be used for linearized massive gravity. As we will see in sect.~\ref{sect:NLFP} (see also \cite{Hinterbichler:2011tt}) the Fierz-Pauli (FP) Lagrangian
\be\label{calLFP1}
{\cal L}_{\rm FP}= \frac{1}{2}
\hmn {\cal E}^{\mu\nu,\rho\sigma}\hrs
- \frac{m^2}{2} (\hmn\hMN -h^2  )\, ,
\ee
(where ${\cal E}^{\mu\nu,\rho\sigma}$ is the  Lichnerowicz operator) is equivalent to a theory with
\be\label{calLFP2}
{\cal L}'=\frac{1}{2}\hmn \(1-\frac{m^2}{\Box}\) {\cal E}^{\mu\nu,\rho\sigma}\hrs
-2m^2N\frac{1}{\Box}\pam\pan(\hMN-\eMN h)\, ,
\ee
where $N$ is an extra field that enters as a Lagrange multiplier. Since
$\pam\pan(\hMN-\eMN h)$ is the linearization of the Ricci scalar, and 
$(1/2)\hmn {\cal E}^{\mu\nu,\rho\sigma}\hrs$ is the quadratic Einstein-Hilbert action, the formulation (\ref{calLFP2}) is manifestly invariant under linearized diffeomorphisms even in the massive case, but is non-local, and again the local formulation (\ref{calLFP1}) can be obtained by imposing a gauge fixing at the level of  the equations of motion.
The theory governed by the first term in \eq{calLFP2},
\be\label{calLnonloc}
{\cal L}_{\rm non-loc}=\frac{1}{2}\hmn \(1-\frac{m^2}{\Box}\) {\cal E}^{\mu\nu,\rho\sigma}\hrs\, ,
\ee
has been first proposed in  the context of the degravitation idea in~\cite{ArkaniHamed:2002fu,Dvali:2007kt}. In ref.~\cite{Dvali:2007kt} it is argued that this theory only  describes two degrees of freedom, corresponding to the states with helicities $\pm 2$ of a massive graviton. A detailed analysis, performed in sect.~\ref{sect:dof}, will show however that it actually describes six radiative degrees of freedom, that make up the five components of a massive spin-2 particle plus a scalar, and  the scalar  is a ghost. Indeed, the constraint imposed by $N$ in \eq{calLFP2} eliminates the ghost, leaving us with the five degrees of freedom of the massive graviton of the linearized FP theory. Furthermore we will show that the scalar, the helicity zero and the helicity $\pm 1$ modes decouple smoothly in the $m\ra 0$ limit, so this theory has no vDVZ discontinuity.

The advantage of this non-local reformulation of the linearized theory will become apparent in sect.~\ref{sect:full}, where we will look for a generally-covariant generalization of the equations of motion derived either from \eq{calLFP2} or from \eq{calLnonloc}. As already observed in \cite{Porrati:2002cp}, the covariantization of \eq{calLFP2} cannot produce a viable theory, since the constraint imposed by $N$ is promoted to the fully covariant constraint $R=0$, which is not present in Einstein theory. Therefore such a theory has a vDVZ discontinuity that persists at the fully non-linear level, and is  ruled out. We will then turn our attention to the covariantization of \eq{calLnonloc}, which in its simplest form is 
\be\label{final1summary}
\Gmn -m^2\(\iBox_g\Gmn\)^{\rm T}=8\pi G\,\Tmn\, ,
\ee
where we used the fact that any symmetric tensor $S_{\mu\nu}$ (here
$S_{\mu\nu}=\iBox_g\Gmn$, where $\Box_g$ is the d'Alembertian in curved space)
can be  decomposed as
$S_{\mu\nu}=S_{\mu\nu}^{\rm T}+(1/2)(\n_{\mu}S_{\nu}+\n_{\nu}S_{\mu})$, where
 $\n^{\mu}S_{\mu\nu}^{\rm T}=0$ (see also \cite{Porrati:2002cp} for a similar approach, applied however to FP theory). The extraction of the transverse part can in principle be performed with non-local operators, which fits well with our general approach.
We will then turn to a discussion 
of the virtues, as well as of the potential problem, of the classical theory defined by \eq{final1summary}.  

The first virtue is that it provides a fully covariant theory of massive gravity, without the need of introducing an external reference metric. Once again, the advantage of having full general covariance  even in a massive theory comes at the price of non-locality. A second important point is that this theory has no vDVZ discontinuity, since the four extra states (the two modes in the scalar sector and the states with helicity $\pm 1$) smoothly decouple in the $m\ra 0$ limit. Thus,  the 
Vainshtein mechanism \cite{Vainshtein:1972sx} is not needed here. Another bonus is that this theory does not have the acausality problem identified in \cite{Deser:2012qx}, since the latter comes from the same constraint that  removes the Boulware-Deser ghost. At the linearized level this is simply the constraint imposed by $N$ in \eq{calLFP2}, which is absent in 
\eq{calLnonloc}. However, these encouraging results seem to come at a disastrous price, namely the existence  of a sixth ghost-like mode, which is already present in the linearized theory
(\ref{calLnonloc}).
We will tackle the ghost issue in sect.~\ref{sect:ghost}, where we will see that this ghost is quite different from the Boulware-Deser ghost that appears
in generic non-linear extensions of FP theory. 
In our case the  ghost has the same mass  $m$ as the  spin-2 graviton, so for
$m\sim H_0$ it is very light. At the same time, in the limit $m\ra 0$ it decouples from the theory and reduces to a non-radiative degree of freedom of GR.
In contrast, the Boulware-Deser ghost is  not
smoothly connected, in the $m\ra 0$ limit, to a harmless non-radiative field, and is not light in a generic background. Rather on the contrary, one usually tries to get rid of it by tuning the parameters of the theory so that its mass goes to infinity. We will see that, as a consequence of the fact that the ghost present in \eq{calLnonloc} decouples in the $m\ra 0$ limit,
the  decay rate of the vacuum due to associated production of ghosts plus positive-energy states is negligible, even over cosmological time-scale. Thus, despite the ghost, the classical theory described by \eq{final1summary} can be perfectly acceptable.
In sect.~\ref{sect:IR} we will examine a variant of \eq{final1summary} of the form
\be\label{final3summary}
\Gmn -m^2\(\frac{1}{\Box_g-\mu^2}\, \Gmn\)^{\rm T}=8\pi G\,\Tmn\, ,
\ee
with $\mu={\cal O}(m^2/\mpl)\ll m$.
This is basically the same as  \eq{final1summary} on field configurations for which
$\Box_g\gg \mu^2$, i.e. for modes that change on a length-scale (or on a time-scale) $L$ such that 
$L\ll \mu^{-1}$, but strongly deviates from it in the far IR, when $L\,\gsim\, \mu^{-1}$. 
The introduction of $\mu$ is particularly interesting since, if we put on the right-hand side a vacuum energy-momentum tensor $\Tmn=-\rho_{\rm vac}\gmn$,  
\eq{final3summary} admits a de~Sitter solution $\Gmn=-\Lambda\gmn$ with
\be
\Lambda=8\pi G\, \frac{\mu^2}{m^2+\mu^2}\,  \rho_{\rm vac}\, .
\ee
Taking now  $\mu\ra  0$ we see that $\Lambda\ra 0$.
This  can be seen as an extreme form of degravitation, in which even in the presence of an  arbitrarily large 
vacuum energy, the  effective cosmological constant $\Lambda ={\cal O}(\mu^2)\ra 0$.
More generally, for finite $\mu$ the vacuum energy $\rho_{\rm vac}$ is degravitated so that the quantity that actually contributes to the observed acceleration of the Universe is
\be
\rho_{\Lambda}=\frac{\mu^2}{m^2+\mu^2}\, \rho_{\rm vac}\, .
\ee
In order to reproduce the observed value $\rho_{\Lambda}={\cal O}(\mpl^2H_0^2)$ from a vacuum energy $\rho_{\rm vac}={\cal O}(\mpl^4)$ we need
$\mu={\cal O}({H_0m/\mpl})$
In particular, for $m={\cal O}(H_0)$, the  vacuum energy that drives the observed acceleration of the Universe is reproduced by a value
\be
\mu={\cal O}\(\frac{m^2}{\mpl}\)\, ,
\ee
which could be naturally generated by gravitational loop corrections.
 We conclude in sect.~\ref{sect:cosmo}, where we show that this non-local theory of massive gravity, specialized to a FRW background, provides a specific model of non-local cosmology.
 Non-local cosmological models have been much studied recently. In particular, in
ref.~\cite{Deser:2007jk} a non-local Friedmann equation has been proposed, obtained 
adding to the Einstein-Hilbert action an extra term of the  form $Rf(\iBox R)$. Its  theoretical structure and cosmological consequences for different choices of the function $f(\iBox R)$ have been discussed in a number of  papers, see e.g.
\cite{Jhingan:2008ym,Koivisto:2008xfa,Koivisto:2008dh,Capozziello:2008gu,Nojiri:2010pw,Bamba:2012ky,Kluson:2011tb,Barvinsky:2011hd,Zhang:2011uv,Elizalde:2011su,Park:2012cp}.
In this approach there is no basic principle that fixes the function $f(\iBox R)$, which  is therefore chosen on purely phenomenological grounds, and can be reconstructed so to fit any given  expansion history~\cite{Deffayet:2009ca,Elizalde:2012ja}. In our case, in contrast, the non-local Friedmann equation follows  from 
\eq{final3summary}, and there is no arbitrary function corresponding to $f$; the  only free parameter is the graviton mass (and $\mu$, which however could in principle be determined in terms of $m$ from the  loop corrections).

Some extra material is discussed in the appendices.
We use the signature $\emn =(-,+,+,+)$ and units $\hbar=c=1$, and we define $\kappa= (32\pi G)^{1/2}$. We use
$\Box=\eMN\pam\pan$ to denote the flat-space d'Alembertian, and $\Box_g$ for the d'Alembertian with respect to a metric $\gmn$.

\section{Non-local formulation of massive electrodynamics}\label{sect:massiveQED}

Before moving to massive gravity let us first discuss  how massive electrodynamics can be written in a gauge-invariant but non-local form. This will be useful to pave the way for the gravitational case, and also has an intrinsic interest. Part of these results have already been presented in \cite{Dvali:2006su,Dvali:2007kt} (see also \cite{Hinterbichler:2011tt} for review). We will however discuss in more detail some technically subtle points involved in the derivation.
We start from the Proca action with an external conserved current $j^{\mu}$
\be\label{1Lemmass}
S=\int d^4x\[ -\frac{1}{4}F_{\mu\nu}F^{\mu\nu}-\frac{1}{2}m_{\g}^2A_{\mu}A^{\mu}
-j_{\mu}A^{\mu}\] \, .
\ee
The equations of motion 
obtained from (\ref{1Lemmass}) are
\be\label{1FAj}
\pam F^{\mu\nu}-m_{\g}^2A^{\nu}=j^{\nu}\, .
\ee
Acting with $\pan$ on both sides and using  $\pan j^{\nu}=0$, \eq{1FAj} gives
\be\label{1mgAmu}
m_{\g}^2\, \pan A^{\nu}=0\, .
\ee
Thus, if $m_{\g}\neq 0$, we get the condition
$\pan A^{\nu}=0$ dynamically, as a consequence of the equation of motion, and we have eliminated one degree of
freedom. Making use of \eq{1mgAmu},
in the vacuum \eq{1FAj} becomes
\be\label{BoxmA0}
(\Box -m_{\g}^2) A^{\mu}=0\, .
\ee
\Eqs{1mgAmu}{BoxmA0} together describe the three degrees of freedom of a massive photon.
In this formulation Lorentz invariance and locality are manifest, while the  $U(1)$ gauge invariance
of the massless theory  is lost, because of the non gauge-invariant term $m_{\g}^2A_{\mu}A^{\mu}$ in the Lagrangian. 

\subsection{Non-local equations of motion}

An equivalent formulation of massive electrodynamics that preserves both Lorentz and gauge invariance by giving up manifest locality can be obtained as follows \cite{Dvali:2006su,Dvali:2007kt}. One begins by performing the ``\Stu trick",    i.e. 
one introduces a scalar  field $\varphi$ and  
replaces
\be
\Am\ra\Am+\frac{1}{m_{\g}}\pam\varphi
\ee
 in the Lagrangian.  Under this replacement $\Fmn$ is unchanged, while the term $\Am j^{\mu}$ only produces a boundary term, since we are assuming that $j^{\mu}$ is conserved. Therefore only the mass term changes, and the new action  is
\be\label{LagStu}
S=\int d^4x\, \[-\frac{1}{4}F_{\mu\nu}F^{\mu\nu}-\frac{1}{2}m_{\g}^2A_{\mu}A^{\mu}
-\frac{1}{2}\pam\varphi\paM\varphi-m_{\g}\AMU\pam\varphi
-j_{\mu}A^{\mu}\]\, .
\ee
By construction  $\Am$ and $\varphi$ only appear in this Lagrangian  in  the combination
$(\Am+m^{-1}_{g}\pam\varphi)$ (apart from  boundary terms, that we will always assume to vanish, setting  appropriate boundary conditions at infinity). Thus the theory is trivially invariant under the local transformation
\be\label{Stulocalsym}
\Am\ra \Am-\pam\theta\, ,\hspace{7mm}\varphi\ra\varphi+m_{\g}\theta\, .
\ee
The equations of motion obtained by taking the variation of the action (\ref{LagStu}) with respect to $\An$ and $\varphi$ are, respectively,
\bees
&&\pam F^{\mu\nu}=m_{\g}^2A^{\nu}+m_{\g}\paN\varphi+j^{\nu}\, ,
\label{QEDeqmotion1}\\
&&\Box\varphi+m_{\g}\pam\AMU=0\, .\label{QEDeqmotion2}
\ees
Of course these equations of motion are invariant under the gauge symmetry (\ref{Stulocalsym}) of the action.
The \Stu field $\varphi$ can then be eliminated from the action by making use of its own equation of motion, that can be written formally as 
\be\label{phiBoxinverse}
\varphi(x)=-m_{\g}\Box^{-1}(\pam\AMU)\, ,
\ee
where, for any integrable  function $f(x)$,
\be
(\iBox f)(x)\equiv \int d^4x'\, G(x;x')f(x')\, ,
\ee
and $G(x;x')$ is a Green's function of the $\Box$ operator, which for the moment we keep generic. Some basic facts about the inversion of the d'Alembertian operator in flat and in curved space are recalled in app.~\ref{app:iBox}.
Substituting \eq{phiBoxinverse} into \eq{QEDeqmotion1} we get
\be\label{eqnonlocAN}
(\Box-m_{\g}^2)\AN=\(1-\frac{m_{\g}^2}{\Box}\)\paN\pam\AMU+j^{\nu}\, .
\ee
Since  \eqs{QEDeqmotion1}{QEDeqmotion2} are invariant under the transformation (\ref{Stulocalsym}), 
the equation of motion (\ref{eqnonlocAN}), which involves only $\Am$,  must be
invariant under the gauge transformation $\Am\ra \Am-\pam\theta$. We can check this immediately observing that, under $\Am\ra \Am-\pam\theta$, the right-hand side of \eq{eqnonlocAN}
changes by a factor
\be
-\(1-\frac{m_{\g}^2}{\Box}\)\paN\Box\theta = -(\Box-m_{\g}^2)\paN\theta\, ,
\ee
which is local, and cancels the change of the left-hand side. Alternatively, we can display the gauge invariance explicitly observing that
\eq{eqnonlocAN} can be 
rewritten as~\cite{Dvali:2007kt,Hinterbichler:2011tt}
\be\label{eqnonlocFMN}
\(1-\frac{m_{\g}^2}{\Box}\)\pan\FMN=j^{\nu}\, .
\ee

\subsection{Non-local action principle}\label{sect:NLaction}

We now wish to find an action whose variation gives \eq{eqnonlocFMN}. It is natural to expect that this is obtained 
performing  the substitution (\ref{phiBoxinverse}) directly into action
(\ref{LagStu}), which gives
\be\label{Lnonloc}
S=-\frac{1}{4}\int d^4x\,\[ \Fmn \(1-\frac{m_{\g}^2}{\Box}\)\FMN
-j_{\mu}A^{\mu}\]\, .
\ee
In fact, the issue is more subtle. Writing \eq{Lnonloc} explicitly, we have
\be\label{Lnonlocexpl}
S=\int d^4x\, \[-\frac{1}{4}\Fmn \FMN-j_{\mu}A^{\mu}\]
+\frac{m_{\g}^2}{4}\int d^4xd^4x' \Fmn(x)G(x;x')\FMN(x')
\, .
\ee
Taking the functional derivatives with respect to $\An$ and to $\pam\An$ we see that the corresponding equation of motion is 
\be\label{eqmotsymmetrized}
\pam\[\FMN (x)-\frac{m_{\g}^2}{2}\int d^4x' [G(x;x')+G(x';x)]\FMN(x')\]=j^{\nu}\, .
\ee
Therefore, as observed also in \cite{Barvinsky:2011rk}, the variational principle automatically symmetrizes the Green's
function, so it
gives back \eq{eqnonlocFMN} only if $G(x;x')=G(x';x)$, i.e.  if 
$\iBox$ is defined  either using the  symmetric combination
$G_+=(1/2)(G_{\rm ret}+G_{\rm adv})$, or the Feynman Green's function $G_F$, see app.~\ref{app:iBox}. There are two possible solutions to this problem:

\vspace{3mm}\noindent
{1.} We indeed use $G_{+}(x;x')$ [or $G_F(x;x')$] in \eq{phiBoxinverse} and therefore in
\eq{eqnonlocFMN}. In this case the non-local action that provides the equations of motion  is indeed given by \eq{Lnonloc}. At first sight, the fact that one uses $G_{+}(x;x')$
or $G_F(x;x')$, which are combinations of $G_{\rm ret}(x;x')$ and $G_{\rm adv}(x;x')$, might seem to pose problems of causality. However, we see from \eq{eqnonlocAN} that the  acausal behavior can be eliminated choosing the gauge $\pam\AMU=0$, and is therefore a gauge artifact that does not affect gauge-invariant observables. This point of view is indeed tenable in a non-local formulation of massive electrodynamics, but is potentially dangerous in a non-local formulation of non-abelian  theories or in the the non-local formulation of fully non-linear massive gravity that we will study in the  sect.~\ref{sect:full},  since non-linear interactions could  communicate the acausal behavior to the physical sector.

\vspace{3mm}\noindent
{2.} Alternatively, we can take the point of view that the classical theory is defined by its equations of motion, while the
action is simply a convenient ``device" that, through a set of well defined rules, allows us to compactly summarize   the equations of motion. We can then take the point of view that the action is
given by \eq{Lnonloc} where $\iBox$ is defined using the symmetric Green's function $G_+$. Then the $\iBox$  operator  is self-adjoint and this allows us to perform standard manipulations such as the integration by parts,
see app.~\ref{app:iBox}. The Euler-Lagrange  equations obtained from this action
are then given by \eq{eqnonlocFMN}, where again $\iBox=\iBox_{+}$. We then add the rule that the physical equations of motion are obtained replacing now
$\iBox_{+}$ with the inverse d'Alembertian computed with the Green's function of our choice, in particular with $\iBox_{\rm ret}$, which ensures causality. 
This is  indeed the 
procedure used in \cite{Soussa:2003vv,Deser:2007jk}, in the context of non-local  gravity theories with a Lagrangian of the form $Rf(\iBox R)$. 

A similar procedure can be used at the quantum level, in the computation of  in-in matrix elements $\langle {\rm in, vac}|\hat{\varphi}|{\rm in,vac}\rangle$
of a quantum field $\hat{\varphi}$, for a Poincar\'e-invariant in-vacuum state in the asymptotic past. In this case one can first work in Euclidean space, computing the Euclidean effective action  in an asymptotically flat space-time. In Euclidean space the Green's function that enters in the $\iBox$ operator is defined imposing vanishing boundary conditions at infinity, and the $\iBox$ operator is unambiguously defined. One can then prove that the nonlocal effective equations for $\langle {\rm in, vac}|\hat{\varphi}|{\rm in,vac}\rangle$
 can be obtained from the Euclidean equations of motion by an  analytic continuation,  
with the prescription that the Euclidean $\iBox$ operator becomes the retarded inverse d'Alembertian $\iBox_{\rm ret}$ in Lorentzian signature \cite{Barvinsky:1987uw}
(see also the discussion in \cite{Barvinsky:2003kg}).

\vspace{3mm}

An equivalent formulation of the latter procedure is obtained  as follows, adapting a construction developed in  \cite{Nojiri:2007uq} in the case of $Rf(\iBox R)$ theories. We
take the action (\ref{Lnonloc}), with a generic $\iBox$, and rewrite it introducing a Lagrange multiplier field $\xi_{\mu\nu}$ as well as an auxiliary field $\psi_{\mu\nu}$, as
\be\label{Semaux}
S=\int d^4x\,\[ -\frac{1}{4}\Fmn \FMN+\frac{m^2_{\g}}{4}\Fmn\psi^{\mu\nu}
+\xi_{\mu\nu}(\Box\psi^{\mu\nu}-\FMN)
-j_{\mu}A^{\mu}\]\, .
\ee
The variation with respect to $\xi_{\mu\nu}$ enforces the constraint
$\psi^{\mu\nu}=\iBox\FMN$, and therefore \eq{Semaux} is formally equivalent to
\eq{Lnonloc}, independently of the Green's function used in the definition of
$\iBox$.
The variations with respect to $\Am$ and $\psi_{\mu\nu}$ give, respectively,
\bees
&&\pam\bigl( \FMN-\frac{m^2_{\g}}{2}\psi^{\mu\nu}+2\xi^{\mu\nu}\bigr)=j^{\nu}\label{Lmultipl1}\\
&&\Box\xi^{\mu\nu}+\frac{m^2_{\g}}{4}\FMN=0\, .
\ees
Substituting $\xi^{\mu\nu}=-(m^2_{\g}/4)\iBox\FMN$ and $\psi^{\mu\nu}=\iBox\FMN$ into
\eq{Lmultipl1} we get \eq{eqnonlocFMN}, independently of the definition of $\iBox$. These manipulations are somewhat formal, since we saw that a proper treatment of the variation
of the action (\ref{Lnonloc}) should rather give \eq{eqmotsymmetrized}. However, in the spirit of point (2.) above, they can be used as a well-defined set of rules that allows us to obtain the equation of motion (\ref{eqnonlocFMN}) from an action.\footnote{At the quantum level, an approach similar in spirit consists in stating that the non-local action is not the fundamental quantity for determining whether the theory is causal. Rather, one must consider the quantum effective action, which is a functional of  the expectation value of the quantum fields. The boundary conditions for the non-local effective action are now fixed by the choice of initial and final quantum states, and can be dealt using the Schwinger-Keldysh technique. A breakdown of causality in the variational equation for the classical fields does not necessarily imply an inconsistency in the computation of 
$\langle{\rm in}|{\rm out}\rangle$ matrix elements, see the discussion in \cite{Barvinsky:2011rk}.}

\vspace{3mm}

In conclusion, the equation of motion (\ref{eqnonlocFMN}) or (with the above qualifications) the action (\ref{Lnonloc}), provide
a formulation of massive electrodynamics in which only the field $\Am$ appears (i.e.  \Stu fields are no longer present), and which is both manifestly Lorentz invariant and gauge invariant. The price that we pay is the lack of manifest locality, since the equation of motion (\ref{eqnonlocAN}) involves the non-local operator $\Box^{-1}$. It should be stressed, however, that the theory {\em is} local, even if not manifestly so, since we have seen that the Lagrangian (\ref{Lnonloc}) is equivalent to the original Proca Lagrangian, which is local.
Observe that we could now  use gauge invariance to fix the Lorentz gauge 
$\pam\AMU=0$. In this way the equation of motion (\ref{eqnonlocAN}) simply becomes
$(\Box-m_{\g}^2)\AN=j^{\nu}$ and the
 non-local term disappears. We therefore get back \eqs{1mgAmu}{BoxmA0} that define Proca theory, except that now the equation $\pam\AMU=0$ emerges as the gauge fixing condition of an underlying gauge theory. In other words, the non-locality only affects pure gauge modes and can be removed by a suitable gauge fixing.

\section{Non-local formulation of Fierz-Pauli  massive gravity}\label{sect:NLFP}

We now consider  FP massive gravity linearized over Minkowski space.
The  action is $S_{\rm FP}+S_{\rm int}$, where  
\be\label{SmassiveG}
\hspace*{-1cm}S_{\rm FP}= \frac{1}{2}\int d^4x\,\[ 
\hmn {\cal E}^{\mu\nu,\rho\sigma}\hrs
- m^2 (\hmn\hMN -h^2  ) \]\, ,
\ee
is
the Fierz-Pauli (FP) action, $\hmn\equiv \gmn-\emn$, and indices are raised and lowered with the flat metric.
The Lichnerowicz operator 
${\cal E}^{\mu\nu,\rho\sigma}$ is defined as
\bees
{\cal E}^{\mu\nu,\rho\sigma}&\equiv&
\frac{1}{2}(\eMR\eNS+\eMS\eNR-2\eMN\eRS) \Box
+ (\eRS\paM\paN+\eMN\paR\paS)\nn\\
&&-\frac{1}{2}\( \eMR\paS\paN+\eNR\paS\paM+\eMS\paR\paN+\eNS\paR\paM\)
\, ,\label{defE}
\ees
where $\Box=\eMN\pam\pan$ is the flat-space d'Alembertian. Therefore
\be\label{Epsh}
{{\cal E}^{\mu\nu,}}_{\rho\sigma}\hRS=\Box\hMN-\eMN\Box h+\eMN\parho\pas\hRS +\paM\paN h-\parho\paN\hMR
-\parho\paM h^{\nu\rho}\, .
\ee
The interaction with the matter energy-momentum tensor is given by
\be\label{Sint}
S_{\rm int}=\frac{\kappa}{2}\int d^4x\, \hmn\TMN \, .
\ee
We take $\TMN$ conserved, so at the linearized level $\pan\TMN=0$.
In order to obtain a gauge-invariant but  non-local formulation of the theory one can introduce a \Stu vector field $\AMU$
through
\be\label{replh}
\hmn\ra \hmn  +\frac{1}{m}(\pam\An+\pan\Am)\, ,
\ee
and then integrate it out using its own equations of motion \cite{Dvali:2006su,Dvali:2007kt,Porrati:2002cp,Hinterbichler:2011tt}. 
By construction, the theory is trivially invariant under the gauge transformation
\be\label{transfhA}
\hmn\ra\hmn  -(\pam\xin+\pan\xim)\, ,\qquad \Am\ra \Am+m\xim\, ,
\ee
that corresponds to  a linearized diffeomorphism. It is often useful to 
perform a further \Stu transformation 
$\Am\ra \Am +(1/m)\pam\varphi$ that introduces a  $U(1)$ symmetry and explicitly extracts  the helicity-0 mode. For the purpose of obtaining the non-local form of massive gravity this step  is not really necessary, so we will only make the replacement (\ref{replh}). 
Then the action becomes
\bees\label{SmassiveGStu}
 S_{\rm FP}+S_{\rm int}&=& \int d^4x\, \[
\frac{1}{2}\hmn {\cal E}^{\mu\nu,\rho\sigma}\hrs
- \frac{m^2}{2} (\hmn\hMN-h^2) -\frac{1}{2}\Fmn\FMN \] \nn\\
&&+\int d^4x\, \[\frac{\kappa}{2}\hmn\TMN+2m\An j^{\nu}\]\, .
\ees
where $\Fmn=\pam\An-\pan\Am$ and
\be\label{defjnu}
j^{\nu}\equiv\pam (\hMN-\eMN h)\, .
\ee
Observe that we could obtain the standard normalization $(-1/4)\Fmn\FMN$ for the kinetic term of $\AMU$ by rescaling $\AMU\ra \AMU/\sqrt{2}$. This would however produce a number of $\sqrt{2}$ factors that would clutter  many subsequent formulas, so we prefer to keep a non-standard normalization for the $\Fmn\FMN$ term.
The  variation with respect to  $\AMU$ gives 
\be\label{eqFhmn}
\pam\FMN=-mj^{\nu}\, .
\ee
Applying $\pan$ to \eq{eqFhmn} we also get the condition 
\be\label{panjnu}
\pan j^{\nu}=0\, .
\ee
One can now eliminate the \Stu field
$\Am$ through its equations of motion. To solve \eq{eqFhmn} we 
separate $\AN$ into its transverse and longitudinal parts, 
\be\label{separTL}
\AN=A^{\nu}_T-\pa^\nu\alpha\, , 
\ee
where $\pan A^{\nu}_T=0$, and we get $\Box A^{\nu}_T=-m j^{\nu}$. Thus, 
the equation of motion of the \Stu field allows us to fix the transverse part to the value
$A^{\nu}_T=-m\Box^{-1}j^{\nu}$, while the longitudinal part remains arbitrary.
This is a peculiarity of the FP mass term, which is such that after the \Stu replacement the kinetic term for the \Stu field $\AMU$ happens to depend only on the $U(1)$-invariant combination
$\Fmn\FMN$. Therefore the longitudinal part, which has the form of a $U(1)$ gauge transformation, remains arbitrary.
Thus, the most general solution of \eq{eqFhmn} is
\cite{Dvali:2006su,Dvali:2007kt}
\be\label{invA1}
\AN=-m\iBox j^{\nu}-\paN\alpha\, ,
\ee
where $\a$ is an arbitrary scalar field. 
Note that, because of \eq{panjnu},
$\pan (\Box^{-1}j^{\nu})=0$ so the term $\Box^{-1}j^{\nu}$ is indeed transverse.
The transformation properties of the field $\alpha$ under linearized diffeomorphisms can be obtained observing that
$\Box\alpha =-\pam\AMU$. Since under linearized diffeomorphisms $\AMU\ra \AMU+m\xiM$, 
we get
\be\label{transalpha}
\Box\alpha\ra\Box\alpha -m\pam\xiM\, .
\ee
Observe that
the transformation property of $(\Box\alpha)/m$,  is the same as that of $h/2$. We then find convenient  to trade $\alpha$ for a new field $N$,
\be\label{defNalpha}
N\equiv 
\frac{h}{2}-\frac{\Box\alpha}{m}
\, ,
\ee
which is   invariant under linearized diffeomorphisms.\footnote{Here our treatment departs  from that in \cite{Dvali:2006su,Dvali:2007kt,Porrati:2002cp}, where $\a$ is fixed to some given value, e.g. $\a=0$. Actually, $\a$ (or, equivalently $N$), is an independent field that will enter the action, and we will see that it  plays the role of a Lagrange multiplier. This gives a more transparent derivation of the constraint associated to FP massive gravity, which otherwise emerges as a consistency condition on the equations of motion.}
Performing the replacement
(\ref{invA1}) in the action (\ref{SmassiveGStu}) and trading $\alpha$ for $N$ we find
\be\label{SmassiveGNonLoc}
S_{\rm FP}+S_{\rm int}= \int d^4x\, \, \[
\frac{1}{2}\hmn \(1-\frac{m^2}{\Box}\) {\cal E}^{\mu\nu,\rho\sigma}\hrs
-2m^2N\frac{1}{\Box}\pam\pan(\hMN-\eMN h)
+\frac{\kappa}{2}\hmn\TMN\] .
\ee
Observe that $N$ enters the action as a Lagrange multiplier.\footnote{\Eq{SmassiveGNonLoc} agrees with the result found with a somewhat different route in
\cite{Hinterbichler:2011tt}, see his eq.~(4.48). Our $\kappa$ corresponds to $2\kappa$  and our $m^2N$ to $N$ in the notation of \cite{Hinterbichler:2011tt}.}   Taking the variation with respect to $N$ we get
\be\label{eqgravNLN}
\pam\pan(\hMN-\eMN h)=0\, ,
\ee 
which, in terms of $j^{\nu}$, can be rewritten as
$\pan j^{\nu}=0$. The variation with respect to $\hmn$ gives 
\be
\label{eqgravNL2}
\(1-\frac{m^2}{\Box}\){{\cal E}^{\mu\nu,}}_{\rho\sigma}\hRS=
-\frac{\kappa}{2} \TMN-2m^2\(\emn-\frac{\paM\paN}{\Box}\)N
\, .
\ee
The field $N$ can be determined algebraically by taking the trace of \eq{eqgravNL2} and using
$\emn {\cal E}^{\mu\nu,\rho\sigma}\hrs=(d-1)\pan j^{\nu}$.
Thus,  upon use of  the equation of motion (\ref{panjnu}), the trace of the left-hand side 
of \eq{eqgravNL2} vanishes, and 
\be\label{NTlin}
N=-\frac{\kappa}{4dm^2}T\, .
\ee
Plugging
\eq{NTlin} into \eq{eqgravNL2} we finally obtain 
\be\label{eqgravNL2noT}
\(1-\frac{m^2}{\Box}\){{\cal E}^{\mu\nu,}}_{\rho\sigma}\hRS=
-\frac{\kappa}{2}\TMN
+\frac{\kappa}{2d}
\(\eMN  -
\frac{\paM\paN}{\Box} \)T
\, .
\ee
Observe that the right-hand side is divergenceless 
(consistently with the linearized Bianchi identity $\pam [{{\cal E}^{\mu\nu,}}_{\rho\sigma}\hRS]=0$) and  traceless. Therefore \eq{eqgravNL2noT} fully summarizes the two equations
(\ref{eqgravNLN}) and (\ref{eqgravNL2}), and provides a non-local formulation of FP massive gravity.

Observe also that one could diagonalize the action (\ref{SmassiveGNonLoc}) with a non-local field redefinition~\cite{Hinterbichler:2011tt}, 
\bees
h'_{\mu\nu}&=&\hmn-\emn\frac{m^2}{\Box-m^2}N\, ,\label{dg1}\\
N'&=&\sqrt{6}\, \frac{m^2}{\Box-m^2}N\, .\label{dg2}
\ees
The action (\ref{SmassiveGNonLoc}) then becomes
\bees\label{SLochatN}
S_{\rm FP}+S_{\rm int}&=& \int d^4x\, \[
\frac{1}{2}h'_{\mu\nu} \(1-\frac{m^2}{\Box}\) {\cal E}^{\mu\nu,\rho\sigma}
h'_{\rho\sigma}
+\frac{1}{2}N'(\Box-m^2)N'\]\nn\\
&&+\frac{\kappa}{2}  \int d^4x\, \(h'_{\mu\nu}\TMN
+\frac{1}{\sqrt{6}}N'T\)
\, .
\ees
However one should be aware that, in general, it is not legitimate  to perform non-local field redefinitions, such as that given in \eqs{dg1}{dg2}, and  use the action written in terms of these non-local fields. The basic point is that operators such as $1/\Box$ or $1/(\Box-m^2)$ are non-local not only in space but even in time, and therefore there is no one-to-one correspondence between the initial conditions on the original fields and on the redefined  fields. This is important in particular when one wants to clearly identify the true dynamical degrees of freedom of the theory. 
In app.~\ref{sect:dyn} we discuss
some examples of the apparent paradoxes in which one can run (even in massless GR) when performing  non-local field redefinitions.

\section{Degrees of freedom of the non-local action}\label{sect:dof}

We now consider  the action 
\be\label{Snon-loc}
S_{\rm non-loc}\equiv \int d^4x\, \,  \frac{1}{2}
{h}_{\mu\nu} \(1-\frac{m^2}{\Box}\) {\cal E}^{\mu\nu,\rho\sigma}
{h}_{\rho\sigma}\, .
\ee
This action was first introduced in the context of the degravitation idea~\cite{ArkaniHamed:2002fu,Dvali:2007kt}, and we have seen that it also enters in FP massive gravity. However, to obtain FP massive gravity, it must be supplemented  by the constraint imposed by the field $N$,
as shown in \eq{SmassiveGNonLoc}. The action $S_{\rm non-loc}$ will be our starting point for the construction of a fully non-linear theory of massive gravity in sect.~\ref{sect:full}, so we will discuss it now in more detail. In particular, we want to understand what degrees of freedom it describes. This is an issue which hides some subtleties, and on which there seems to be some confusion in the literature. 

The propagating degrees of freedom of a theory can  be read from the propagator.
In GR one starts from the quadratic Einstein-Hilbert  action
\be
S_{\rm EH}^{(2)}=\frac{1}{2} \int d^4x\, \hmn {\cal E}^{\mu\nu,\rho\sigma}\hrs\, .
\ee
To obtain the propagator one must add a gauge fixing term. A convenient choice is 
\be\label{1Sgf}
S_{\rm gf}=-\int d^4x\, \(\paN\bhmn \) (\parho\bar{h}^{\rho\mu})\, .
\ee
where $\bhmn=\hmn -(1/2)h\emn$. Then
\be
S_{\rm EH}^{(2)}+S_{\rm gf}=\int d^4x\, \[ - \frac{1}{2}\parho\hmn\paR\hMN 
+\frac{1}{4}\paM h\pam h 
\]\, .
\ee
Inverting this quadratic form one finds the propagator of massless gravitons,
\be\label{Dgrav}
\tilde{D}^{\mu\nu \rho\s}(k)=\frac{1}{2}\( \eMR\eNS +\eMS\eNR-\eMN\eRS \)
\(\frac{-i}{k^2-i\eps}\)\, ,
\ee
where the $i\eps$ prescription selects the Feynman
propagator.
Consider now the  non-local action   (\ref{Snon-loc}). This action is gauge invariant, so we  need again a gauge fixing. We find convenient to use as gauge-fixing term
\be
S_{\rm gf}=  -\frac{1}{\xi}\int d^4x\, \, \(\paN\bhmn \) 
\(1-\frac{m^2}{\Box}\) 
 (\parho\bar{h}^{\rho\mu})\, ,
 \ee
and use the  gauge $\xi=1$. After some integration by parts we get
\bees
S_{\rm non-loc}+S_{\rm gf}&=&\int d^4x\, \[
 - \frac{1}{2}\parho  h_{\mu\nu}\(1-\frac{m^2}{\Box}\) \paR {h}^{\mu\nu} 
+\frac{1}{4}\paM h \(1-\frac{m^2}{\Box}\) \pam h\]\nn\\
&=&
\frac{1}{2}\int d^4x\, {h}^{\mu\nu}
A_{\mu\nu\rho\s}(\Box-m^2){h}^{\rho\sigma}\, ,
\ees
where
\be
A_{\mu\nu\rho\s} =\frac{1}{2}
\( \emr\ens +\ems\enr-\emn\ers \)\, .
\ee
Observe that in  gauge $\xi=1$ the non-local terms in $S_{\rm non-loc}+S_{\rm gf}$ cancel. This gives again an example of the interplay between gauge invariance and non-locality. We can write the action in a gauge-invariant form at the price of non-locality, or in a local form at the price of fixing a suitable gauge.
The  propagator in this gauge is obtained inverting this
quadratic form, which gives 
\be\label{propSnonloc}
\tilde{D}^{\mu\nu \rho\s}(k)=\frac{1}{2}\( \eMR\eNS +\eMS\eNR-\eMN\eRS \)
\(\frac{-i}{k^2+m^2-i\eps}\)\, .
\ee
Thus, the tensor structure is the same as in massless GR, and the only change is in the overall factor $-i/(k^2-i\eps)$, which becomes $-i/(k^2+m^2-i\eps)$. Observe that in the theory defined by $S_{\rm non-loc}$ there is no vDVZ discontinuity, and in the $m\ra 0$ limit the propagator (\ref{propSnonloc}) smoothly reduces  to the massless propagator (\ref{Dgrav}).

Naively one might think that, since the tensor structure of the propagator in massless GR and in $S_{\rm non-loc}$ are the same, the radiative degrees of freedom are the same, too. If this were the   case, 
$S_{\rm non-loc}$ would only contain two massive states with helicities $\pm 2$. However, this reasoning  is incorrect.
This can be first  illustrated comparing massless and massive electrodynamics.
The propagator of the massless photon is
\be\label{A17}
\tilde{D}^{\mu\nu}(k)=\frac{-i}{k^2-i\eps} \[\eMN -(1-\xi)\frac{k^{\mu}k^{\nu}}{k^2}\]\, ,
\ee
For conserved  currents,  in momentum space we have $k^{\mu}\tilde{j}_{\mu}(k)=0$, so the term proportional to $k^{\mu}k^{\nu}$ in the propagator does not contribute, and the saturated propagator is
\bees\label{jj}
\tilde{j}_{\mu}(-k)\tilde{D}^{\mu\nu}(k)\tilde{j}_{\nu}(k)&=&
\frac{-i}{k^2-i\eps}\,
\emn\tilde{j}^{\mu}(-k)\tilde{j}^{\nu}(k)\\
&=&\frac{-i}{k^2-i\eps}\[ 
-\tilde{j}^{0}(-k)\tilde{j}^{0}(k) +\tilde{j}^{3}(-k)\tilde{j}^{3}(k)
+\sum_{i=1,2}\tilde{j}^{i}(-k)\tilde{j}^{i}(k)\]\, .\nn
\ees
In the massless case this tensor structure, proportional to $\eMN$, describes the exchange of only the states with helicities $\pm 1$. In fact, for an on-shell photon we can write  $k_{\mu}=\omega (-1,0,0,1)$, and then current conservation  implies $\tilde{j}^{0}(k)=\tilde{j}^{3}(k)$. Thus the first two term in \eq{jj} cancel, and
the interaction mediated by an on-shell massless photon is proportional to 
\be
\sum_{i=1,2}\tilde{j}^{i}(-k)\tilde{j}^{i}(k)=
  \tilde{j}^{+}(-k) \tilde{j}^{-}(k)+ 
\tilde{j}^{-}(-k) \tilde{j}^{+}(k)
\, ,
\ee
where $\tilde{j}^{\pm}=(\tilde{j}^{1}\pm i\tilde{j}^{2})/\sqrt{2}$.
This shows that the interaction of on-shell massless photons only involves the operators
$\tilde{j}^{\pm}$, which have helicities $\pm 1$, and therefore is of the form
$\tilde{A}_{+}(-k)\tilde{j}_{-}(k)+\tilde{A}_{-}(-k)\tilde{j}_{+}(k)$, where
$\tilde{A}_{\pm}$ are fields with helicities $\pm 1$. 

The propagator of the massive photon is instead
\be\label{1propmasskk}
\tilde{D}^{\mu\nu}(k)=\frac{-i}{k^2+m_{\g}^2-i\eps}\, 
\(\eMN +\frac{k^{\mu}k^{\nu}}{m_{\g}^2} \)\, .
\ee
Again, for a conserved current the term $k^{\mu}k^{\nu}/m_{\g}^2$ does not contribute so the massive photon propagator can be taken to be
\be\label{1propmasskk2}
\tilde{D}^{\mu\nu}(k)=\frac{-i}{k^2+m_{\g}^2-i\eps}\, 
\eMN \, ,
\ee
so its tensor structure is effectively given simply by $\eMN$,  just as for the massless propagator. However, now
$k_{\mu}=(-\omega,0,0,k)$ with $\omega=(k^2+m_{\g}^2)^{1/2}$, and
the current conservation equation $k_{\mu}\tilde{j}^{\mu}=0$  gives
$\tilde{j}^{0}(k)=(k/\omega)\tilde{j}^{3}(k)$. Then 
the terms
$-\tilde{j}^{0}(-k)\tilde{j}^{0}(k) +\tilde{j}^{3}(-k)\tilde{j}^{3}(k)$ no longer cancel. Rather, now 
\be\label{jj2}
\tilde{j}_{\mu}(-k)\tilde{D}^{\mu\nu}(k)\tilde{j}_{\nu}(k)=
\frac{-i}{k^2-i\eps}\,\[  \tilde{j}^{+}(-k) \tilde{j}^{-}(k)+ 
\tilde{j}^{-}(-k) \tilde{j}^{+}(k)
+\frac{m_{\gamma}^2}{\omega^2} \tilde{j}^{3}(-k)\tilde{j}^{3}(k)\]\, ,
\ee
showing that there is an extra term that describes the coupling of the longitudinal polarization. Thus, even if the tensor structure of
the propagator (\ref{1propmasskk2}) is the same as that of the massless propagator, still it describes two transverse and one longitudinal degrees of freedom, as of course  should be for a massive photon. Observe also that, for $m_{\gamma}\ra 0$, the longitudinal mode smoothly decouples.

The situation is completely analogous when comparing the  propagator of massless GR with that of $S_{\rm non-loc}$. In the massless case the tensor structure in \eq{Dgrav} reflects the fact that a massless graviton only has the helicities $\pm 2$. In fact, in momentum space
energy-momentum conservation reads $k_{\mu}\tilde{T}^{\mu\nu}(k)=0$. For on-shell massless gravitons we can write again $k_{\mu}=\omega (-1,0,0,1)$ , and 
energy-momentum conservation becomes
\be\label{T00T33}
\tilde{T}^{0\nu}(k)=\tilde{T}^{3\nu}(k)\, .
\ee
We can now compute explicitly the saturated propagator $\tilde{T}_{\mu\nu}(-k)
\tilde{D}^{\mu\nu \rho\s}(k)\tilde{T}_{\rho\sigma}(k)$, 
and eliminate all occurrences of 
$\tilde{T}^{0\nu}(k)$ using  \eq{T00T33}. Then   one finds that the terms involving a spatial index $i=3$ cancel, and  
\be\label{A23}
\tilde{T}_{\mu\nu}(-k)
\tilde{D}^{\mu\nu \rho\s}(k)\tilde{T}_{\rho\sigma}(k)\,
 =\tilde{T}_{-2}(-k)\frac{-i}{k^2-i\eps}\,\tilde{T}_{+2}(k)
 +\tilde{T}_{+2}(-k)\frac{-i}{k^2-i\eps}\,\tilde{T}_{-2}(k)
\ee
where 
\be
\tilde{T}_{\pm 2}= \frac{1}{2}
\( \tilde{T}_{11}-\tilde{T}_{22}\mp 2i \tilde{T}_{12}\)\, .
\ee
Under rotations by an angle $\theta$ around the $z$ axis  the combinations $\tilde{T}_{\pm 2}$ transform as $\tilde{T}_{\pm 2}\ra \exp\{\pm 2i\theta\}\tilde{T}_{\pm 2}$, and are therefore eigenstates of the helicity with eigenvalue $\pm 2$.
This shows that this propagator describes  a massless particle with helicities $\pm 2$.\footnote{
Indeed, the graviton propagator (\ref{Dgrav}) can be found without performing explicitly the inversion of the quadratic form in the action, simply observing that it must be symmetric in $(\mu,\nu)$ and in
$(\rho,\sigma)$. Thus, it can only depend on the 
combinations $(\emr\ens +\ems\enr)$ and $\emn\ers$
(apart for term involving $k_{\mu}k_{\nu}$, $k_{\rho}k_{\sigma}$, etc.
that gives zero when contracted with the energy-momentum tensor. The particular choice of gauge fixing gauge used in \eq{1Sgf} actually sets these terms to zero). Requiring that
the combination $(\emr\ens +\ems\enr+a\emn\ers)$ selects the helicity-2 part of the energy momentum tensor and using $k_{\mu}\tilde{T}^{\mu\nu}(k)=0$, with $k_{\mu}=\omega (-1,0,0,1)$,
fixes $a=-1$ \cite{Boulware:1973my}.}

In contrast, for massive gravitons one must again write $k_{\mu}=(-\omega,0,0,k)$ with $\omega=(k^2+m^2)^{1/2}$, and the conservation equation $k_{\mu}\tilde{T}^{\mu\nu}(k)=0$ no longer reduces the saturated propagator to a form that only involves the helicity-2 operators. Rather, we now have 
$0=k_{\mu}\tilde{T}^{\mu\nu}(k)=-\omega \tilde{T}^{0\nu}(k)+k\tilde{T}^{3\nu}(k)$,
so 
\be\label{T0nu}
\tilde{T}^{0\nu}(k)=(k/\omega)\tilde{T}^{3\nu}(k)\, . 
\ee
We use this relation to eliminate all occurrences of $\tilde{T}^{0\nu}$, and  we introduce \bees
\tilde{T}_{\pm 1}&=&\tilde{T}_{13}\mp i \tilde{T}_{23}\, ,\\
\tilde{T}_0&=&3\, \tilde{T}_{33}\, .
\ees
The five quantities $\tilde{T}_q(k)$, with $q=-2,\ldots,2$ are helicity eigenstates with eigenvalue $q$, and their normalizations have been chosen for later convenience.   The four-dimensional trace
$\tilde{T}(k)=\eMN\tilde{T}_{\mu\nu}(k)$ is instead a Lorentz scalar. From
\eq{T0nu} we have
$\tilde{T}^{00}(k)=(k/\omega)\tilde{T}^{30}(k)$ and
$\tilde{T}^{30}(k)=\tilde{T}^{03}(k)=(k/\omega)\tilde{T}^{33}(k)$, so
$\tilde{T}^{00}(k)=(k/\omega)^2\tilde{T}^{33}(k)$. This gives
\be\label{T112233}
\tilde{T}(k)=\eMN\tilde{T}_{\mu\nu}(k)=\tilde{T}_{11}(k)+\tilde{T}_{22}(k)
+\frac{m^2}{\omega^2}\tilde{T}_{33}(k)\, .
\ee
Eliminating $\tilde{T}^{0\nu}(k)$ through \eq{T0nu} and
trading the six quantities $\tilde{T}_{ij}(k)$ for the five components of a spin-2 operator
$\tilde{T}_{q}(k)$, $q=-2,\ldots, 2$ plus the scalar $T$, we get
\bees\label{TmT}
\tilde{T}_{\mu\nu}(-k)\tilde{D}^{\mu\nu \rho\s}(k)\tilde{T}_{\rho\sigma}(k)\,
 &=&\sum_{q=-2,2}\tilde{T}_{-q}(-k)\frac{-i}{k^2+m^2-i\eps}\,\tilde{T}_{q}(k)\nn\\
&& +\frac{m^2}{\omega^2}\sum_{q=-1,1}\tilde{T}_{-q}(-k)\frac{-i}{k^2+m^2-i\eps}\,\tilde{T}_{q}(k)\\
&&+ \frac{m^2}{6\omega^2}\, \, \(T_0(-k),T(-k)\) \frac{-i D}{k^2+m^2-i\eps}
\(\begin{array}{cc}
T_0(k)\\T(k)
\end{array}\)
\, ,\nn
\ees
where the matrix $D$ is given by
\be
D=\(\begin{array}{cc}
m^2/\omega^2 & -1\\
-1 &0\\
\end{array}\)\, .
\ee
The eigenvalues of $D$ are $\lambda_{\pm}=\eps\pm \sqrt{1+\eps^2}$ where
$\eps=m^2/(2\omega^2)$, so $\lambda_+>0$ and $\lambda_{-}<0$, corresponding to a particle with the good sign in the propagator and a ghost, respectively. The  eigenvectors are the combinations $t_{\pm}=T_0+\lambda_{\mp}T$, which to lowest order in $\eps$ reduce to $T_0\mp T$. The  fields that diagonalize the propagator in the scalar sector are therefore  the corresponding combinations of the helicity-0 mode and of the scalar field.
We see that the propagator of $S_{\rm non-loc}$ describes 6 dynamical fields: besides the expected massive states with helicities $q=\pm 2$, there are two states with helicities $q=\pm 1$, a state with helicity $q=0$ (which, together, form the states of a spin-2 massive particle), and a scalar field. In the limit $m\ra 0$ the contribution of the helicities $\pm 1$ goes smoothly to zero, because it is multiplied by an overall factor $m^2/\omega^2$.  The same happens in the scalar sector.\footnote{This shows   that  the statement in \cite{Dvali:2007kt} that the action $S_{\rm non-loc}$ describes  only
the  states with helicity $\pm 2$, once the  helicities $0$ and $\pm 1$ have been integrated out, is   incorrect.
The integration over the \Stu field $\Am$ should not be confused with the integration over the helicity-0 and helicity-1 modes. When we perform the \Stu replacement (\ref{replh}) we are formally increasing the number of fields in the theory, and this increase is compensated by the appearance of a gauge symmetry. Thus, 
after the replacement $\hmn\ra \hmn  +(1/m)(\pam\An+\pan\Am)$, the field
$\hmn$ still contains its helicity-0, helicity-1 and helicity-2 states and, furthermore, we have introduced extra helicity-0 and helicity-1 states associated to $\Am$. When we eliminate the latter (either integrating out $\Am$, or for instance just choosing the gauge $\Am=0$) we still remain with the helicity-0,  helicity-1 and helicity-2 states associated with $\hmn$, and  the action $S_{\rm non-loc}$   still contains
two scalars, two states with helicities $\pm 1$ and two states with helicities $\pm 2$.}

This counting of degrees of freedom is confirmed observing that we have been able to rewrite FP massive gravity in the form (\ref{SmassiveGNonLoc}). Here the field $N$ enters as a Lagrange multiplier, so it is not dynamical and it  enforces the single constraint (\ref{eqgravNLN}). This constraint therefore removes one scalar degree of freedom from the six described by $S_{\rm non-loc}$, and we remain with five degrees of freedom, in agreement with the fact that
\eq{SmassiveGNonLoc} is just a rewriting of linearized FP 
massive gravity.

Of course, the above counting of degrees of freedom  can also be derived from the invariance  of the action (\ref{Snon-loc})  
under linearized diffeomorphisms
${h}_{\mu\nu}\ra{h}_{\mu\nu}  -(\pam\xin+\pan\xim)$. 
Using this invariance, out of the 10 components of $\hmn$ we can  eliminate four (and only four) degrees of freedom from  the action
(\ref{Snon-loc}). 
Following the steps that in the massless case lead to the TT gauge
we can in fact  use the four functions $\xiM$ to fix the Lorentz gauge
\be \label{Lorgauhbmn}
\pan\bhMN =0\, ,
\ee
where $\bhMN =({h}^{\mu\nu}-(1/2)\eMN {h})$.  Under gauge transformations
\be
\pan\bhMN\ra \pan\bhMN -\Box\xiM\, , 
\ee
so fixing the Lorentz gauge
leaves a residual gauge invariance parametrized by four functions $\xiM$ that satisfy
$\Box\xiM=0$.  
In linearized {\em massless} gravity, after fixing the Lorentz gauge, the metric satisfies 
$\Box\hMN=0$, so the residual gauge invariance can  be used to set to zero four more components of $\hmn$, namely one transverse vector and two scalars. Thus, out of the original ten components of $\hmn$, four are eliminated by \eq{Lorgauhbmn} and four more by the residual gauge invariance, and we remain with the two degrees of freedom of a massless graviton, corresponding to the helicities $\pm 2$.
In the massive case the residual gauge symmetry cannot be used to eliminate further degrees of freedom. Indeed,
the equation of motion derived  from \eq{Snon-loc}
is
\be\label{eqSnonloc}
\(1-\frac{m^2}{\Box}\){{\cal E}^{\mu\nu,}}_{\rho\sigma}{h}^{\rho\sigma}=0
\, .
\ee
Observe that 
\be\label{Ehbar}
{{\cal E}^{\mu\nu,}}_{\rho\sigma}{h}^{\rho\sigma}=\Box \bhMN-
\paM\parho\bar{h}^{\rho\nu}-\paN\parho\bar{h}^{\rho\mu}+\eMN
\parho\pas\bar{h}^{\rho\sigma}\, .
\ee
Thus, in the Lorentz gauge we have ${{\cal E}^{\mu\nu,}}_{\rho\sigma}{h}^{\rho\sigma}=\Box \bhMN$ and \eq{eqSnonloc} becomes  local, and  is just a massive KG, $(\Box-m^2)\bhmn=0$.
Contracting with $\emn$ we also have $(\Box-m^2) {h}=0$, 
and therefore in the end, after fixing the gauge (\ref{Lorgauhbmn}), the equation of motion for $\hmn$ becomes
\be
(\Box-m^2){h}^{\mu\nu}=0\, .
\ee
Using functions $\xiM$ which are constrained to obey $\Box\xiM=0$ we cannot eliminate components of $\hMN$ that satisfy
$\Box\hMN\neq 0$. Thus,  we find again the action (\ref{Snon-loc}) describes six 
degrees of freedom, which corresponds  to the five components of a massive spin-2 particle, plus a Lorentz scalar.\footnote{
This situation is completely analogous to what happens  in the non-local formulation of massive electrodynamics discussed in sect.~\ref{sect:massiveQED}.
The action  (\ref{Lnonloc}) describes  three (rather than two) radiative degrees of freedom. This is evident from the fact that it is just a rewriting of the original Proca theory, which describes a massive photon and therefore three radiative degrees of freedom. 
One might be puzzled by the fact that 
\eq{Lnonloc} describes three propagating degrees of freedom because this action is gauge-invariant, and we are used to the fact that a $U(1)$ gauge invariance removes two degrees of freedom. However, again the point is that  the single function $\theta$ that parametrizes the gauge transformation $\Am\ra\Am-\pam\theta$ allows us to eliminate two degrees of freedom only in the massless case. This can be seen for instance using the $U(1)$ symmetry to fix the Lorentz gauge $\pam\AMU=0$. This leaves us with the freedom of performing  a residual gauge transformation $\Am\ra\Am-\pam\theta$ with $\Box\theta=0$. In massless electrodynamics,
in the Lorentz gauge  the equation of motion in vacuum $\pam\FMN=0$ becomes
$\Box\AN=0$ and we can use the residual gauge freedom to set $A^0=0$,  and then 
$\pam\AMU=0$ becomes $\n\cdot{\bf A}=0$. We have therefore reached 
the radiation gauge $A^0=0, 
\n\cdot{\bf A}=0$. Thus, when $m_{\g}=0$, the single function $\theta$  can be used to eliminate both $A^0$ and the longitudinal component of the photon.
In contrast, when $m_{\g}\neq 0$, after fixing the Lorentz $\pam\AMU=0$,  we remain with the equation $(\Box-m_{\g}^2)\AN=0$, and the residual gauge invariance parametrized by a function $\theta$ with $\Box\theta=0$ cannot be used to eliminate a further degree of freedom.}
A further insight into the structure of the non-local action $S_{\rm non-loc}$ can be gained by introducing non-local variables in terms of which the action takes a local form. This provides a rather elegant formulation, which is discussed in app.~\ref{app:nonlocfields}.

\section{A covariant  fully non-linear theory of massive gravity}\label{sect:full}

We now show how to construct a viable covariant, fully non-linear theory of massive gravity using this non-local formulation.

\subsection{Covariantization of FP theory}

Consider first the FP action, in the form (\ref{SmassiveGNonLoc}).
To perform the 
covariantization
we begin by observing that, linearizing around flat space, $\gmn=\emn+\hmn$,
we have $R=R^{(1)}+{\cal O}(h^2)$, where
\be
R^{(1)}=\pam\pan (\hMN-\eMN h)\, .
\ee
Thus, the simplest covariant generalization of the term $N \iBox\pam\pan (\hMN-\eMN h)$ in the action (\ref{SmassiveGNonLoc}) is just $N\iBox R$, where the field $N$ is promoted to be a scalar under full diffeomorphisms, and the covariantization of
\eq{eqgravNLN} is simply $R=0$. This condition was already found, with a different route,
in \cite{Porrati:2002cp}, where it was also correctly observed that it provides a discontinuity with the massless theory, since in  GR we rather have  $R=-8\pi G T$.
At the linearized level this is just  the vDVZ discontinuity and we see that in this covariantization it persists at the level of the fully non-linear theory. Such a covariantization therefore necessarily leads to a theory in conflict with the experiment, even assuming that it gives a consistent theory.
Of course the covariantization procedure is not unique, and one could rather replace
$N\pam\pan (\hMN-\eMN h)$ with $N[ R+{\cal O}(R_{\mu\nu\rho\sigma}^2)]$, which still has the correct linearized limit. However,  this would still give rise to a constraint that is not present in GR, and that reduces to $R=0$ at low curvatures. We will therefore turn our attention to
$S_{\rm non-loc}$, and construct a covariant generalization of this theory, rather than of FP theory.

\subsection{Covariantization of $S_{\rm non-loc}$}
We find convenient to work at the level of the equations of motion, so we look for a covariantization of \eq{eqSnonloc} including also the source term,
\be\label{eqSnonloc2}
\(1-\frac{m^2}{\Box}\){{\cal E}^{\mu\nu,}}_{\rho\sigma}{h}^{\rho\sigma}=
-16\pi G\TMN
\, ,
\ee
where we have  rescaled $\hmn\ra \kappa\hmn$, so that now in \eq{eqSnonloc2}
$\hmn$ is dimensionless, and we used $\kappa^2=32\pi G$. In this section we continue to use the notation $\Box$ for the flat-space d'Alembertian, while we  denote by $\Box_g$ the d'Alembertian computed with respect to a generic metric $\gmn$.
The covariant generalization of the left-hand side can be found observing that
the linearization of the Einstein tensor over Minkowski is given by
$\Gmn =\Gmn^{(1)}+{\cal O}(h^2)$, with
\be
\Gmn^{(1)} =-\frac{1}{2}{\cal E}_{\mu\nu,\rho\sigma}\hRS \, .
\ee
Thus, a generally covariant expression that reduces to the right-hand side of \eq{eqSnonloc2} is $-2(1-m^2/\Box_g)\GMN$. Of course, this is not the only possible expression that has the correct linearized limit. However, we must further require that the correct generalization of the left-hand side is a covariantly conserved tensor,  in order to be consistent with $\n_{\mu}\TMN=0$. We then proceed as in \cite{Porrati:2002cp}, and observe that any symmetric tensor $S_{\mu\nu}$ can be  decomposed as
\be\label{splitSmn}
S_{\mu\nu}=S_{\mu\nu}^{\rm T}+\frac{1}{2}(\n_{\mu}S_{\nu}+\n_{\nu}S_{\mu})\, , 
\ee
where
$\n^{\mu}S_{\mu\nu}^{\rm T}=0$. One could further decompose
$S_{\mu\nu}^{\rm T}$ into
a transverse-traceless part $S_{\mu\nu}^{\rm TT}$ and the trace part, and similarly $S_{\mu}=S_{\mu}^{\rm T}+\n_{\mu} \Sigma$, where $\n^{\mu}S_{\mu\nu}^{\rm T}=0$ and $\n^{\mu}S_{\mu}^{\rm T}=0$.
The various components can be extracted explicitly with the use of non-local operators, and in flat space  the explicit expressions are given  in 
\eqst{decomphmn}{defhath}. For 
a general metric the explicit expressions are more complicated, basically because the covariant derivative does not commute with $\Box_g$. 

In terms of this decomposition,  a natural covariantization of \eq{eqSnonloc2} is
\be\label{final0}
\[ \(1-\frac{m^2}{\Box_{g}}\)\Gmn\]^{\rm T}=8\pi G\,\Tmn\, ,
\ee
i.e.
\be\label{final1}
\Gmn -m^2\(\iBox_g\Gmn\)^{\rm T}=8\pi G\,\Tmn\, ,
\ee
where the superscript T denotes the operation of taking the transverse part. By construction the divergence of the left-hand side vanishes, 
so we still have $\n^{\mu}\Tmn=0$.\footnote{Of course, one can always add  (the transverse part of) quantities quadratic in the Riemann tensor to the left-hand side, since these do not affect the linearized limit.  By dimensional reasons, these terms must be suppressed by the inverse of a mass squared. If they are suppressed by $1/\mpl^2$, these terms are irrelevant much below the Planck scale. However, having at our disposal both $m$ and $\mpl$, one can also in principle  write a theory where such terms are suppressed by  $1/\Lambda^2$ with $\Lambda^n=m^{n-1}\mpl$ for some $n$ (i.e., one of the scales that appear in the \Stu description  of the local formulation of massive gravity). In this case, \eq{final1} should be regarded as the IR limit of this more general class of theories, and these extra terms could be important for the UV completion of the theory.}
The classical theory defined by \eq{final1} 
is a covariant, fully non-linear theory of massive gravity defined without introducing a reference metric. As discussed in the introduction, this is conceptually quite satisfying, since the introduction of a reference metric basically means that we have a different definition of the massive theory for  every background of the massless theory. Furthermore this theory has no vDVZ discontinuity since its propagator, given by \eq{propSnonloc}, reduces smoothly to the GR propagator as $m\ra 0$.

Note also that, if we use the retarded  Green's function in the $\iBox_g$ operators that appear in \eq{final1}, the theory is non-local  but only involves an integration over the past light cone, and therefore preserves causality. Observe also that the  problems of superluminal propagation discussed in \cite{Deser:2012qx} are generated by the same constraint that eliminates the ghost in FP theory.\footnote{However, as mentioned in the introduction, the observation of   ref.~\cite{Deser:2012qx} does not yet imply 
the loss of causality in dRGT, since in the attempt of constructing closed time-like curves one might be forced to leave the domain of validity of the effective field theory \cite{Burrage:2011cr}, so in dRGT the causality problem is rather postponed to the UV completion. See \cite{Babichev:2007dw} for a review of the issue.}
Since this constraint is absent in
the theory defined by $S_{\rm non-loc}$, this particular example of superluminality is also absent. Thus  causality problems, at least in the form identified to date in non-linear extensions of FP massive gravity, are not present (although a detailed analysis is needed to study whether other forms of superluminality might emerge in some specific background).

\section{The ghost problem}\label{sect:ghost}

\begin{figure}[ht]
\begin{center}
\begin{minipage}{1.\linewidth}
\centering
\includegraphics[width=0.2\textwidth]{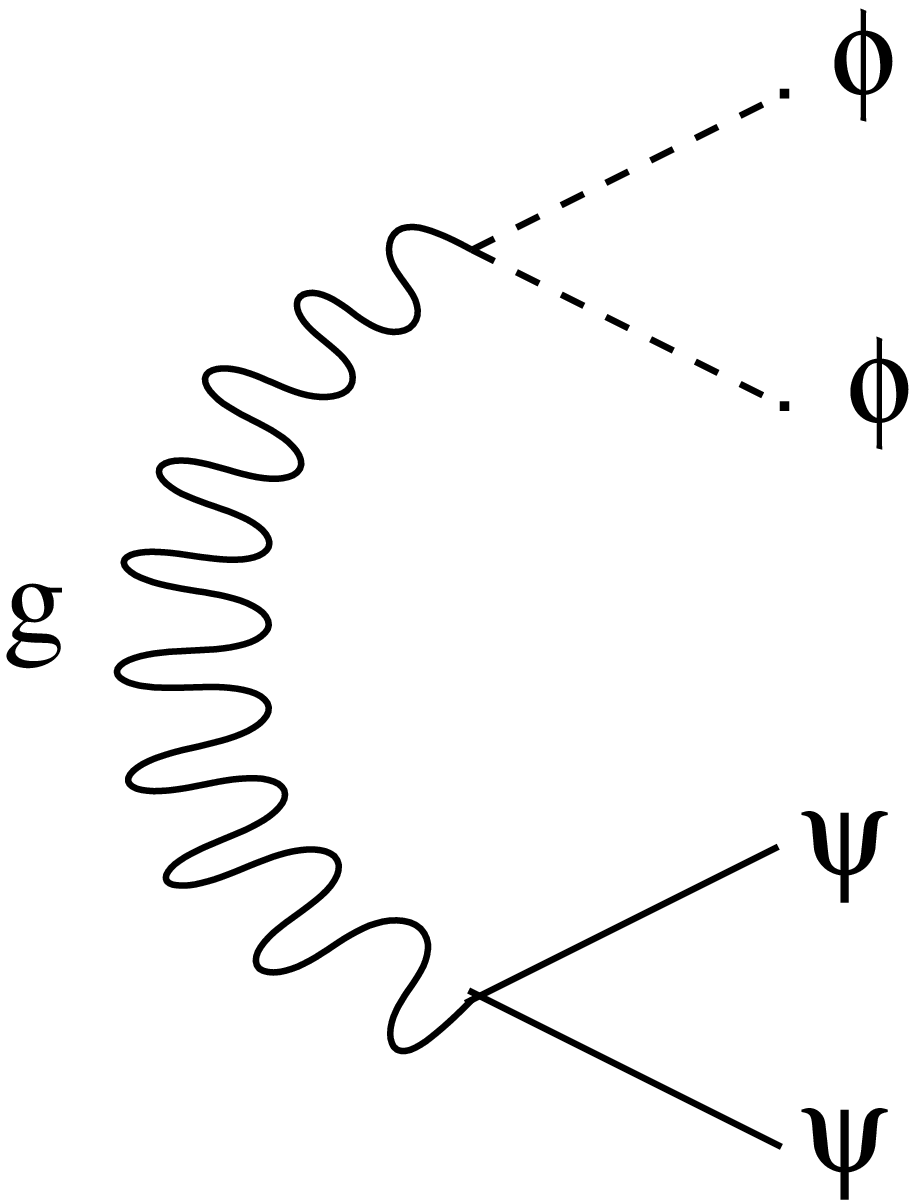}\hspace{3cm}\quad
\includegraphics[width=0.2\textwidth]{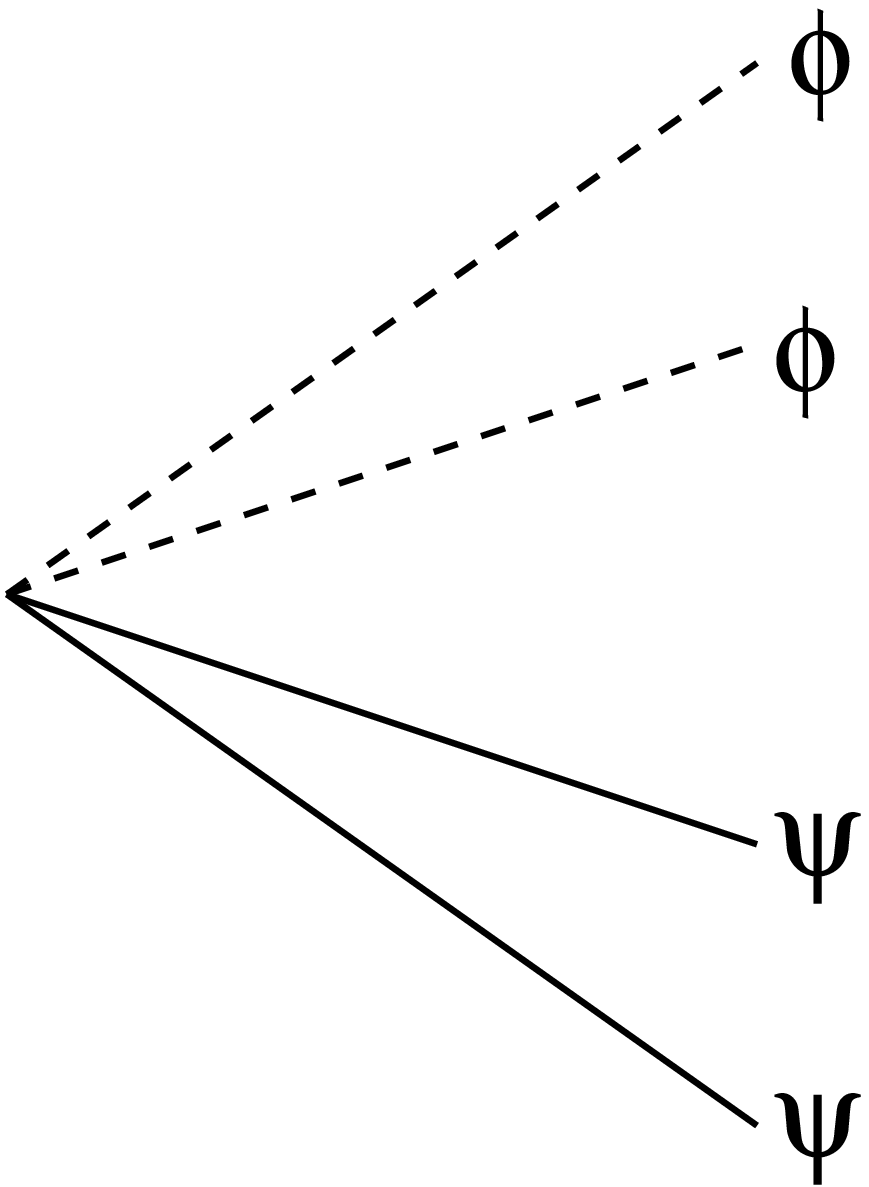}\hspace{3cm}
\end{minipage}
\caption{\label{fig:vacdecay} Left: the Feynman graph describing 
${\rm (vacuum)}\ra \psi\psi\phi\phi$. The wavy line denotes any of the six states described by
$\hmn$.
Right: the same process, mediated directly by the four-point vertex.}
\end{center}
\end{figure}

As we found in sect.~\ref{sect:dof}, one degree of freedom in $S_{\rm non-loc}$ is a ghost.
At first this seems to doom the theory (\ref{final1}) to failure.  In particular, one might fear that the vacuum decays  quickly through associated production of positive-energy massive gravitons and negative-energy ghosts, see e.g. the discussion
in  \cite{Carroll:2003st,Cline:2003gs,Kaplan:2005rr,Garriga:2012pk}.
In our case, as discussed in sect.~\ref{sect:dof}, in the scalar sector we have a healthy state  $\psi$ and a ghost state $\phi$, which are linear combinations of the helicity-0 component of the massive spin-2 graviton and of the scalar degree of freedom.  In the covariantization of $S_{\rm non-loc}$ we have for instance a trilinear gravitational vertex 
proportional to $h\pa h\pa h$ 
(where $h$ denotes symbolically the 5 components of the massive graviton plus the scalar), which induces 
processes such as  ${\rm (vacuum)}\ra \psi\psi\phi\phi$ through diagrams such as that on the left of   Fig.~\ref{fig:vacdecay}. The four-point interaction $h h\pa h\pa h$ is instead responsible for the diagram on the right of   Fig.~\ref{fig:vacdecay}. 

Observe however that in our case the mass of the ghost has the same value $m$ as the mass of the spin-2 graviton. This result is a consequence of the structure of the propagator in 
$S_{\rm non-loc}$, and is protected by the diffeomorphism invariance of the covariantization of $S_{\rm non-loc}$. For cosmological applications the graviton mass $m$  must  be very small, of the order of the present value of the Hubble parameter $H_0$. As a consequence, our ghost is extremely light, too. Furthermore, and quite crucially, we have seen that in this theory there is no vDVZ discontinuity, and the extra scalar and vector polarizations decouple smoothly in the 
limit $m\ra 0$, see \eq{TmT}. Indeed, we prove in app.~\ref{app:NLgra} that in the $m\ra 0$ limit the ghost smoothly goes into a non-radiative degree of freedom of GR.
We therefore expect that, for $m={\cal O}(H_0)$,  instabilities associated to the ghost only develop at most on  cosmological time-scales. In this sense, the existence of the ghost could even be welcome, since a phase of  accelerated expansion of the Universe can  be seen as  an instability. 
We will now put this physical intuition on a more formal basis  by estimating  the probability of ghost-induced vacuum decay.\footnote{It should be stressed that the ghost that appears in $S_{\rm non-loc}$  and in its covariantization  is  quite different from   the Boulware-Deser ghost that appears
in non-tuned non-linear extension of FP theory. The Boulware-Deser ghost  on a generic background  gets a mass fixed by the scales of the background \cite{Creminelli:2005qk}, and  is not
smoothly connected, in the $m\ra 0$ limit, to a harmless non-radiative field. Rather on the contrary, one tries to get rid of it by making it very heavy. Indeed,  at the linearized level the FP tuning sends the ghost mass to infinity. The problem is that this procedure is in general unstable against the introduction of non-linearities, and  when expanding the theory over a generic background a finite ghost mass reappears. The tuning of the potential in the dRGT theory~\cite{deRham:2010ik,deRham:2010kj} is indeed performed so to send again the mass of the ghost above the cutoff scale of the effective theory. }

\subsection{Ghost-induced vacuum decay rate}

Before examining the computation  in the non-local formulation of massive gravity, let us 
consider a simpler theory with action
\be\label{toy}
S=\int d^4x\, \[ \frac{1}{2}(-\pam\psi\paM\psi -m_{\psi}^2\psi^2)+
\frac{1}{2}(\pam\phi\paM\phi -m_{\phi}^2\phi^2)+\frac{\lambda}{4} \phi^2\psi^2\]\, .
\ee
With our ``mostly plus" signature $\psi$ is a healthy scalar while $\phi$ is a ghost. We first  
recall how the computation is performed in this simpler case \cite{Cline:2003gs,Kaplan:2005rr,Garriga:2012pk}. We consider for definiteness the graph on the right of Fig.~\ref{fig:vacdecay} and
we denote by $k_1,k_2$ the momenta of the normal particles, and by $p_1,p_2$ the momenta of the ghosts. For the normal particles the energies  are positive, $k_1^0>0, k_2^0>0$, while for the ghosts
$p_1^0<0, p_2^0<0$. We also introduce the notation $\omega_i=k_i^0$, and $E_i=-p_i^0$
($i=1,2$). The amplitude associated to the graph on the right-hand side of Fig.~\ref{fig:vacdecay}
is given by
\be
i{\cal M}_{fi}=i\lambda (2\pi)^4 \d^{(4)}(p_1+p_2+k_1+k_2)\, .
\ee
Regularizing the theory in a spatial volume $V$ and a time interval $T$,
and using 
\be
\left|(2\pi)^4 \d^{(4)}(p)\right|^2=VT(2\pi)^4\d^{(4)}(p)\, , 
\ee
the differential probability of vacuum decay, $dw$, is given by 
\be
dw=\lambda^2 (2\pi)^4 \d^{(4)}(p_1+p_2+k_1+k_2) VT
\frac{1}{2!}\frac{d^3p_1}{(2\pi)^3 2E_1}\frac{d^3p_2}{(2\pi)^3 2E_2}
\frac{1}{2!}\frac{d^3k_1}{(2\pi)^3 2\omega_1}\frac{d^3k_2}{(2\pi)^3 2\omega_2}\, ,
\ee
(where the $1/2!$ are the factors for identical particles). This is the probability that the decay happens anywhere in space and at any time, so the decay probability per unit volume and unit time is
\be
\Gamma =\frac{w}{VT}=\frac{\lambda^2}{2!2!} \int 
\frac{d^3p_1}{(2\pi)^3 2E_1}\frac{d^3p_2}{(2\pi)^3 2E_2}
\frac{d^3k_1}{(2\pi)^3 2\omega_1}\frac{d^3k_2}{(2\pi)^3 2\omega_2}
 (2\pi)^4\d^{(4)}(p_1+p_2+k_1+k_2) \, .
\ee
This can be conveniently manipulated introducing the identities
\be
1=\int d^4P\, \d^{(4)}(P-p_1-p_2)\, ,\qquad
1=\int d^4K\, \d^{(4)}(K-k_1-k_2)\, .
\ee
Then
\bees
\Gamma &=&\frac{\lambda^2}{(2\pi)^4} \int d^4Pd^4K\,   \d^{(4)}(P+K)
\[\frac{1}{2!}\int \frac{d^3p_1}{(2\pi)^3 2E_1}\frac{d^3p_2}{(2\pi)^3 2E_2}
\, (2\pi)^4\d^{(4)}(P-p_1-p_2) \]\nn\\
&&\times \[ \frac{1}{2!}\int \frac{d^3k_1}{(2\pi)^3 2\omega_1}\frac{d^3k_2}{(2\pi)^3 2\omega_2}
\, (2\pi)^4\d^{(4)}(K-k_1-k_2) \]\nn\\
&=&\frac{\lambda^2}{(2\pi)^4} \int d^4Pd^4K\,   \d^{(4)}(P+K)
\Phi_{\phi}^{(2)}(-P^2)\Phi_{\psi}^{(2)}(-K^2)\nn\\
&=&\frac{\lambda^2}{(2\pi)^4} \int d^4P\, 
\Phi_{\phi}^{(2)}(-P^2)\Phi_{\psi}^{(2)}(-P^2)\, ,\label{Gam1}
\ees
where $\Phi^{(2)}_{\phi}$  is the two-body  phase space for two identical particles of mass $m_{\phi}$, and  depends on $P$ only through the Lorentz scalars $-P^2$, and similarly for 
$\Phi_{\psi}^{(2)}(-P^2)$ (observe that, with our signature, 
$-P^2$ and $-K^2$ are positive). 
The two-body phase space for two identical particles of mass $m$ is
\be
\Phi^{(2)}(s)=\theta(s-4m^2) \frac{1}{16\pi}\, \sqrt{1-\frac{4m^2}{s}}\, ,
\ee
and goes to a constant in the large $s$ limit. Thus, the integral over $d^4P$ diverges. 
To better understand this divergence
we can further manipulate \eq{Gam1} inserting the identity in the form
\be
\int_0^{\infty} ds\, \d(s+P^2)=1\,,
\ee
where $P^2=-(P^0)^2+{\bf P}^2$. Then
\bees
\Gamma&=&\frac{\lambda^2}{(2\pi)^4}\int d^4P\, \Phi_{\phi}^{(2)}(-P^2)\Phi_{\psi}^{(2)}(-P^2) \int_0^{\infty} ds\, \d(s+P^2)\nn\\
&=&\frac{\lambda^2}{(2\pi)^4}\int_0^{\infty} ds\, \Phi_{\phi}^{(2)}(s)\Phi_{\psi}^{(2)}(s) \int dP^0d^3{\bf P}\, \d\(s-(P^0)^2+{\bf P}^2\)\nn\\
&=&\frac{\lambda^2}{(2\pi)^4}\int_0^{\infty} ds\, \Phi_{\phi}^{(2)}(s)\Phi_{\psi}^{(2)}(s) \int dP^0d^3{\bf P}\, \frac{1}{2P^0}\d\(P^0-\sqrt{{\bf P}^2+s}\)\nn\\
&=&\frac{\lambda^2}{(2\pi)^4}\int_0^{\infty} ds\, \Phi_{\phi}^{(2)}(s)\Phi_{\psi}^{(2)}(s) \int  \frac{d^3{\bf P}}{2\sqrt{{\bf P}^2+s}}\, .\label{Icutoff}
\ees
Both the  integral over $s$ and that  over the three-momentum ${\bf P}$ diverge. 
Putting a cutoff $|{\bf P}|<\Lambda$ as well as $s<\Lambda^2$ and using
the asymptotic form $\Phi^{(2)}(s)\simeq 1/(16\pi)$ one obtains
\be\label{estiGam}
\Gamma\sim \lambda^2\, \(\frac{\Lambda}{8\pi}\)^4\, .
\ee
Observe that,
despite the fact that   $d^3{\bf P}/(2\sqrt{{\bf P}^2+s})$ is a Lorentz-invariant measure,  it cannot be regularized preserving Lorentz invariance. Putting a cutoff  $|{\bf P}|<\Lambda$ breaks Lorentz invariance, and therefore such a cutoff should come from new Lorentz-violating physics~\cite{Cline:2003gs,Kaplan:2005rr}.\footnote{One might consider the possibility of regularizing the original integral over $d^4P$ by
rotating into Euclidean space and putting a Lorentz-invariant cutoff $\Lambda$ over the modulus of the Euclidean momentum, $(P_E^0)^2+{\bf P}^2<\Lambda^2$.
However, the usual Wick rotation is an operation that is performed on the  integrals that enter in the loop corrections to the amplitudes, and is justified by the analyticity properties of the amplitudes. Here the Euclidean rotation would  rather be performed on the phase space integrals, and it is not obvious that  it makes any sense.}
It has been shown in \cite{Garriga:2012pk} that, in the presence of non-local interactions that become soft in the UV (which is indeed our case), the momentum integral can become convergent, and only the integral over $s$ need a regularization. A cutoff $s<\Lambda^2$ does not spoil Lorentz invariance,  so in this case Lorentz-violating physics is no longer required. The whole issue of Lorentz violations becomes   irrelevant if we put the cutoff $\Lambda$ at the Planck scale, since in any case we expect that beyond $\mpl$ a Lorentz-invariant and local QFT description will no longer  be appropriate. However, for $\Lambda\sim\mpl$,
\eq{estiGam} provides a decay rates of about one event per Planck volume per Planck time so (unless one choses a ridiculously small value of $\lambda$) the vacuum decay is basically instantaneous.  This means that, in the theory (\ref{toy}), with no new physics until  $\Lambda\sim \mpl$, the ghost instability is  a fatal one.

Consider next what happens if 
a ghost interacts only gravitationally, but still in a local theory. Then the situation is somewhat different because in the coupling enters $G=1/\mpl^2$, and there is a natural UV cutoff
$\Lambda\sim \mpl$, beyond which one might reasonably assume that string theory or any other UV completion takes over and softens the gravitational interaction.
In normal GR the structure of the Lagrangian is schematically of the form
\be 
{\cal L}\sim \frac{1}{G} \[ \pa h\pa h +( h \pa h\pa h)
+( h h\pa h\pa h)
\]\, .
\ee
After rescaling $h\ra G^{1/2}h$ to get a canonically normalized kinetic term, we get a Lagrangian of the form
\be
{\cal L}\sim \pa h\pa h+ G^{1/2}\(  h \pa h\pa h\)
+G( h h\pa h\pa h)\, ,
\ee
and therefore the four-graviton vertex gives a contribution proportional to $GE^2$, where $E$ is an energy scale of the process. In a local theory of massive gravity in which $\hmn$ contains the five degrees of freedom of the massive graviton plus a scalar ghost, in order of magnitude the computation will therefore be the same as above, with the replacement 
\be
\lambda\ra \lambda_{\rm eff}(s)\sim  Gs\, .
\ee
Thus 
\be\label{GammaL6}
\Gamma \sim \frac{1}{(2\pi)^4}\, \frac{1}{(16\pi)^2}\, \int^{\Lambda^2} ds (Gs)^2 
\int^{\Lambda}\, 2\pi|{\bf P}|d|{\bf P}|
\sim   \(\frac{\Lambda}{8\pi}\)^4\,\frac{\Lambda^4}{\mpl^4}\,.
\ee
A contribution of the same order comes from the graphs on the left-hand side of
Fig.~\ref{fig:vacdecay}, due to a factor $(G^{1/2}s)$ from each vertex and a factor $\sim 1/s$ from the graviton propagator. Again,
for $\Lambda\sim \mpl$, \eq{GammaL6}  gives a production rate of order one event per Planck time, per Planck volume, and  we therefore get a catastrophic decay that immediately destabilizes the vacuum. This is indeed what happens if we consider the vacuum decay induced by the ghost that is present in the local formulation of non-linear generalizations of FP massive gravity, in agreement with the discussion in 
\cite{Gabadadze:2003jq}.

Consider now  the vacuum decay rate in the 
non-local theory of massive gravity that we are proposing.
To obtain an order-of-magnitude  estimate  of the vacuum decay rate  we can argue as follows. 
The  interaction involving the ghost can only come from the non-local sector (given 
that for $m=0$ the ghost decouples and reduces to a non-radiative degree of freedom). 
Thus, compared to the standard gravitational case, the interaction is softened by a factor $m^2/\Box$. This factor reflects the fact that the ghost matches smoothly a non-radiative degree of freedom in the massless limit, and there is no vDVZ discontinuity in our theory.
Independently of the details of the action, this should  contribute an extra  factor ${\cal O}(m^2/s)$ to the amplitude, and hence a factor $m^4/s^2$ to the probability. Thus, in order of magnitude, the vacuum decay rate in the non-local theory can be estimated as\footnote{Observe that, for the purpose of this order-of-magnitude estimate, it is irrelevant whether the term $m^2/\Box$ contributes to the integral over $s$ with a factor $m^4/s^2$, or to the integral over $|{\bf P}|$ with a factor $m^4/|{\bf P}|^4$. In the latter case it would render UV finite the integral over $d^3P$, as in \cite{Garriga:2012pk}.
If the boost integral is not regularized by the $m^2/\Box$ term, than we need to rely on physics beyond the Planck scale for its regularization. However, this does not necessarily mean that physics beyond the Planck scale must violate Lorentz invariance in order to regularize the vacuum decay rate (indeed, Lorentz violations even at the Planck scale  are severely constrained \cite{Myers:2003fd}). Rather, a UV completion such as string theory could provide an effective  non-locality, that regularizes the boost integral similarly to what happens in
\cite{Garriga:2012pk}. 
}
\be\label{GammadE}
\Gamma \sim \frac{1}{(2\pi)^4}\, \frac{1}{(16\pi)^2}\, \int^{\Lambda^2} ds (Gs)^2 \, \frac{m^4}{s^2}
\int^{\Lambda}\, 2\pi|{\bf P}|d|{\bf P}|
\sim \(\frac{m}{8\pi}\)^4\, \frac{\Lambda^4}{\mpl^4}\, .
\ee
We see that, even for $\Lambda\simeq \mpl$, the rate does not exceed a value of order
$[m/(8\pi)]^4$. Taking $m\sim 8\pi H_0$, for the production  of a $\psi\psi\phi\phi$ final state out of the vacuum,
this gives a rate of  one event in a volume equal to the present Hubble volume $H_0^{-3}$, over the whole age of the universe $t\sim H_0^{-1}$. 
Such a  rate is totally  irrelevant. To get a sense for it, 
consider the energy density $\rho_{\psi}$ in $\psi$ particles produced per unit time by this process (of course, an equal and opposite energy density is produced in ghosts, and the total energy density is conserved). This is obtained multiplying the rate $\Gamma$ (number of events per unit time per unit volume) by the energy carried by each event. Since the integral in
\eq{GammadE} is dominated by the UV cutoff region, setting $\Lambda\sim \mpl$ this is simply
$\dot{\rho}_{\psi}\sim (m/8\pi)^4\mpl$. We can compare it with the evolution of the energy density of the Universe due to the standard cosmological expansion; writing 
$\rho_{\rm tot}\sim H^2\mpl^2$, we have
$\dot{\rho}_{\rm tot}\sim H\dot{H}\mpl^2\sim H^3\mpl^2$. Thus, \be
\frac{\dot{\rho}_{\psi}}{\dot{\rho}_{\rm tot}}\sim \(\frac{m}{8\pi H}\)^4 
\, \frac{H}{\mpl}\, .
\ee
This quantity is an increasing function of time, and for 
$m= {\cal O}(8\pi H_0)$ 
even at the present epoch it is minuscule, of order $H_0/\mpl\sim 10^{-60}$.
Indeed, to make it of order one at the present epoch, we would need a mass $m$ parametrically larger than $H_0$, $m\sim 8\pi H_0 (\mpl/H_0)^{1/4}\sim 10^{16} H_0$ (i.e.
$m^{-1}\sim 0.1$~au). For smaller values of $m$, and in particular for the values $m\sim H_0$  of cosmological interest, the process of ghost-induced vacuum decay  is irrelevant.

Observe that, to get this result, it was crucial that the ghost interaction is softened by the term $m^2/\Box$, giving an extra  factor $m^2/E^2$ in the amplitude and finally an extra factor 
$m^4/\Lambda^4$ in the rate. Without this factor, the decay rate would have rather been 
given by \eq{GammaL6}. The crucial difference with the Boulware-Deser ghost that appears in generic non-linear extensions of FP gravity,   can therefore be traced back to  the fact that the non-local theory (\ref{final1}) has no vDVZ discontinuity, and all extra degrees of freedom, including the ghost, decouple in the $m\ra 0$ limit, as shown by \eq{TmT}.

\section{de~Sitter solutions and degravitation}\label{sect:IR}

It is   interesting to observe that \eq{final1} does not admit  exact de~Sitter  solutions
$\Gmn^{\rm dS}=-\Lambda\gmn$ (observe that, with our signature,  de~Sitter corresponds to $\Lambda>0$). In fact, in this case
$(\iBox_g \Gmn)^T=
-\Lambda(\iBox_g \gmn)^T=-\Lambda\iBox_g \gmn$ (because $\gmn$ is already transverse).
Since  $\Box_g\gmn=0$, we have
$\iBox_g\gmn=\infty$. Thus, in de~Sitter,  the term $(\iBox_g \Gmn)^T$
diverges.  To understand this problem, let us first introduce a new parameter $\mu$, and consider the equation
\be\label{final5}
\Gmn -m^2\(\frac{1}{\Box_g-\mu^2}\, \Gmn\)^{\rm T}=8\pi G\,\Tmn\, .
\ee
We have chosen the sign of
$\mu^2$ in 
\eq{final5} so that $\mu^2>0$ corresponds to a non-tachyonic mass term.
We can think of $\mu$ as a regulator that will  eventually be sent to zero, but  it is in fact quite interesting to consider the theory with finite $\mu$. In particular,  will see that it is especially interesting to take a value $\mu={\cal O}(m^2/\mpl)$.  In this case $\mu\ll  m$, and
\eq{final1} is now seen as an approximation which is only  valid for modes with typical spatial or temporal variations much smaller than $1/\mu$, so that on these modes $\Box_g\gg\mu^2$.

\Eq{final5} admits de~Sitter solutions, which show interesting degravitation properties. Consider an energy-momentum tensor of the form 
$\Tmn=-\rho_{\rm vac}\gmn=(\rho_{\rm vac},-a^2\rho_{\rm vac}\d_{ij})$, so 
$p_{\rm vac}=-\rho_{\rm vac}$, and look for a solution
$\Gmn=\Gmn^{\rm dS}=-\Lambda\gmn$. Now this solution exists, and $\Lambda$ is fixed by
\be
\Lambda=8\pi G\, \frac{\mu^2}{m^2+\mu^2}\,  \rho_{\rm vac}\, .
\ee
Taking now  $\mu\ra  0$ at fixed $m$ we see that $\Lambda\ra 0$.
This  can be seen as an extreme form of degravitation in which, even in the presence of an  arbitrarily large 
vacuum energy, the  effective cosmological constant $\Lambda ={\cal O}(\mu^2)\ra 0$.
More generally, for finite $\mu$ the vacuum energy $\rho_{\rm vac}$ is degravitated so that the quantity that actually contributes to the observed acceleration of the Universe is
\be\label{rhoLrhov}
\rho_{\Lambda}=\frac{\mu^2}{m^2+\mu^2}\, \rho_{\rm vac}\, .
\ee
In order to reproduce the observed value $\rho_{\Lambda}={\cal O}(\mpl^2H_0^2)$ from a vacuum energy $\rho_{\rm vac}={\cal O}(\mpl^4)$ we need
\be
\mu={\cal O}\(\frac{H_0m}{\mpl}\)\, .
\ee
In particular, if $m={\cal O}(H_0)$, we need
\be
\mu={\cal O}\(\frac{m^2}{\mpl}\)\, .
\ee
Such a value, which would   provide a  natural solution for the cosmological constant problem,  is just of the  size that can be  expected from  gravitational loop corrections: if fact, in the 
 bubble graph giving the one-loop graviton self-energy diagram, which provides the correction  $\delta m^2$ to the graviton mass, each of the two trilinear vertices gives a factor $\sqrt{G}\sim 1/\mpl$, so $\delta m^2\sim 1/\mpl^2$ (and the same for the one-loop correction to the propagator  involving a single four-graviton vertex). In the limit $m\ra 0$ we must have $\delta m^2\ra 0$ since in this case mass renormalization is protected by general covariance, so it is natural to expect $\mu^2\sim\delta m^2\sim m^4/\mpl^2$.

The fact that the only exact de~Sitter solution has the above value of $\Lambda$ 
is not a problem for inflation because, if we take $m$ of order of the present Hubble rate $H_0$, in the early Universe \eq{final5} admits  quasi-de~Sitter solutions which are practically indistinguishable from the usual slow-roll inflationary solutions. Indeed, in a spatially uniform time-dependent background 
$\Box\sim d^2/dt^2={\cal O}(\omega^2)$, where $\omega$ is the characteristic frequency of variation of the background, and the non-local term in \eq{final5} is negligible if
$\omega^2\gg m^2$ (which also implies $\omega^2\gg \mu^2$).  For a FRW metric with Hubble parameter $H(t)$, in particular, the characteristic frequency is $\omega=|\dot{H}|/H$. In terms of the slow-roll parameter $\eps=-\dot{H}/H^2$ 
the condition $\omega\gg m$ reads $\eps\gg m/H(t)$. If we take $m={\cal O}(H_0)$,
in the early Universe when inflation takes place, $m/H(t)$ is a ridiculously   small number
(e.g. of order $10^{-57}$ for GUT-scale inflation), much smaller than the typical values, say $\eps= {\cal O}(10^{-2})$, of the slow-roll parameter. More generally, for $m={\cal O}(H_0)$, the non-local term  is irrelevant in the early Universe and only 
becomes important  in the recent cosmological epoch. In other words, 
\eq{final5} works as a high-pass filter that degravitates all sources with a typical frequency $\omega$ smaller that $\mu$, or a typical length-scales larger than $\mu^{-1}$ (which, for 
$m={\cal O}(H_0)$ and $\mu={\cal O}(m^2/\mpl)$, is parametrically larger than the horizon size). In particular, an exactly constant vacuum energy is totally degravitated. However, the  source term due to an inflaton  field slowly rolling into a potential, even in the ``slow"-roll regime,  still evolves with a characteristic frequency which is huge compared to $\mu$, and is not affected at all by the non-local terms.\footnote{Observe also that, linearizing \eq{final5}, we  get a propagator $\tilde{D}(p)=-i (p^2+\mu^2)/[p^2(p^2+m^2+\mu^2)]$.  Beside the pole at $p^2=-(m^2+\mu^2)\simeq -m^2$ we therefore now have an extra pole at $p^2=0$. Its residue is however proportional to $\mu^2/(m^2+\mu^2)$ which,
for $\mu\sim m^2/\mpl$ and $m\sim H_0$,  is  order $H_0^2/\mpl^2$. These extra state are therefore totally decoupled on subhorizon and even on horizon scales, and their only role is to degravitate the vacuum energy to the value (\ref{rhoLrhov}) rather than down to zero.}

Finally it is interesting to observe that, if we rather take a tachyonic value $\mu^2=-m^2$,
\eq{final5} becomes 
\be\label{final3}
\Gmn -m^2\(\frac{1}{\Box_g+m^2}\, \Gmn\)^{\rm T}=8\pi G\,\Tmn\, ,
\ee
which is equivalent to
\be\label{final4}
\(\frac{\Box_g}{\Box_g+m^2}\, \Gmn\)^{\rm T}=8\pi G\,\Tmn\, .
\ee
In this case,  degravitation is lost.
However, this equation is  interesting because it admits a family of self-inflationary solution, i.e. de~Sitter solutions
$\Gmn=-\Lambda\gmn$ with arbitrary $\Lambda$, in the absence of any external source,
$\Tmn=0$. Again, as long as $|\dot{H}/H|\gg m$, the non-local term in \eq{final3} is negligible. Thus, taking $m$ of order $H_0$, the early Universe cosmology is unaffected, and in particular we have the standard radiation dominated and matter dominated (MD) era. However, when $H(t)$ drops to values comparable to $H_0$, even in a theory without a explicit  cosmological constant term in the action, we expect that the MD solution will be attracted by one of these self-inflationary de~Sitter solutions.

\section{Non-local cosmology from  massive gravity}\label{sect:cosmo}

The non-local modification of GR that we are proposing induces a non-local modification of the Friedmann equation.  
To derive the equations of non-local cosmology we specialize
\eq{final5} to a flat FRW metric. We use coordinates $(t,\vx)$ where  $t$ is cosmic time, so $\gmn=(-1,a^2(t)\d_{ij})$ and $\Tmn=(\rho,a^2p\d_{ij})$, and we work for generality in $d$ spatial dimensions. The evolution of the scale factor is determined by the
non-local generalization of the Friedmann equation, i.e. by the $(00)$ component of \eq{final5}, together with energy-momentum conservation, which is ensured by the fact that the left-hand side of \eq{final5} is transverse.
Introducing
\be
S_{\mu\nu}\equiv \frac{1}{\Box_g-\mu^2}\Gmn\, 
\ee
and splitting
$S_{\mu\nu}$ as in \eq{splitSmn}, we can rewrite \eq{final5} as the coupled system of equations
\bees
\Gmn -m^2S_{\mu\nu}^{\rm T}&=&8\pi G\,\Tmn\, ,\label{Fin1}\\
(\Box_g-\mu^2)S_{\mu\nu}&=&\Gmn\, .\label{Fin2}
\ees
To extract the transverse part from $S_{\mu\nu}$ we  take the divergence of \eq{splitSmn}. Then $S_{\mu\nu}^{\rm T}$ drops and we get
\be\label{fixSnu}
\n^{\mu}(\n_{\mu}S_{\nu}+\n_{\nu}S_{\mu})=2\n^{\mu}S_{\mu\nu}\, .
\ee
These four equations determine the four components of $S_{\mu}$ in terms of 
$S_{\mu\nu}$. Then $S_{\mu\nu}^{\rm T}$ is obtained in terms of $S_{\mu\nu}$ by
\be\label{fixST}
S_{\mu\nu}^{\rm T}=S_{\mu\nu}-\frac{1}{2}(\n_{\mu}S_{\nu}+\n_{\nu}S_{\mu})\, .
\ee
On a generic background it can be non-trivial to find the solution of  \eq{fixSnu}. However, in  FRW the solution can be obtained  very simply  observing that in this case 
there is no preferred spatial direction,  so the only possible solution of \eq{fixSnu} for the spatial vector $S^i$ is $S^i=0$. \Eq{fixSnu} with $\nu=0$ then suffices to determine $S_0$,
\bees
S_0&=&-\frac{1}{\pa_0^2+dH\pa_0-dH^2}\, \n_{\mu}S^{\mu}_{0}\nn\\
&=&-\frac{1}{\pa_0^2+dH\pa_0-dH^2}\,\( \dot{u}+dHu-Hv\)\, ,
\ees
where $H(t)=\dot{a}/a$, the dot is the derivative with respect to cosmic time $t$, and we have introduced the variables
\be
u(t)=S^0_0(t)\, ,\qquad v(t)=S^i_i(t)\, ,
\ee
(where sum over $i=\{1,\ldots d\}$ is understood).
From \eq{fixST} we  have $S_{00}^{\rm T}=S_{00}-\n_0S_0=
S_{00}-\dot{S}_0$ since, in the coordinates $(t,\vx)$ we have  $\Gamma^{\mu}_{00}=0$.
Therefore 
\bees
(S^0_0)^{\rm T}&=&S^0_0+\pa_0S_0\nn\\
&=&u-\pa_0\frac{1}{\pa_0^2+dH\pa_0-dH^2}\,\( \dot{u}+dHu-Hv\)\, ,
\ees
and of course $H=H(t)$ so we must be careful with the ordering of 
$\pa_0$ and $[\pa_0^2+dH\pa_0-dH^2]^{-1}$. 
This allows us to write \eq{Fin1} in terms of $H(t),u(t)$ and $v(t)$,
We now turn to \eq{Fin2}.
Evaluating the $\Box$ operator on a rank-$(1,1)$ tensor we get
\bees
\Box S^0_0&=&-\ddot{S}^0_0-dH\dot{S}^0_0+2d H^2S^0_0-2H^2S^{i}_i\, ,\\
\Box S^i_i&=&-\ddot{S}^i_i-dH\dot{S}^i_i-2dH^2S^{0}_0+2 H^2S^i_i\, .
\ees
Finally, in $d$ spatial dimensions for the Einstein tensor in a FRW background we have
\be
G^0_0=-\frac{d(d-1)}{2}H^2\, ,\qquad
G^i_i=-d(d-1)\dot{H}-\frac{d^2(d-1)}{2}H^2\, .
\ee
Putting these results together we get a system of three coupled equations for the three variables $\{H(t),u(t),v(t)\}$: the $(00)$ component of \eq{Fin1} gives the modified Friedmann equation
\be\label{uv0}
\frac{d(d-1)}{2}H^2+m^2\[ u-\pa_0\frac{1}{\pa_0^2+dH\pa_0-dH^2}\,\( \dot{u}+dHu-Hv\)\]=8\pi G\rho\, ,
\ee
while the $(00)$ and $(ii)$ components of \eq{Fin2} give, respectively,
\bees
\ddot{u}+\mu^2u+dH\dot{u}-2d H^2u+2H^2v&=&\frac{d(d-1)}{2}H^2\label{uv1}\\
\ddot{v}+\mu^2v+dH\dot{v}+2dH^2u-2 H^2v&=&
d(d-1)\dot{H}+\frac{d^2(d-1)}{2}H^2\, .\label{uv2}
\ees
\Eqs{uv1}{uv2} can be decoupled introducing
\be
U=u+v\, ,\qquad V=u-\frac{1}{d}v\, .
\ee
Observe that $U=S^0_0+ S^i_i=\gMN S_{\mu\nu}$. Then we get the system of equations
\be
\frac{d(d-1)}{2}H^2+\frac{m^2}{d+1}\[ U+dV-\pa_0\frac{1}{\pa_0^2+dH\pa_0-dH^2}\,\( \dot{U}+d\dot{V}+d(d+1)HV\)\]=8\pi G\rho\, , \label{UV0}\ee
\bees
&&\ddot{U}+\mu^2U+dH\dot{U}=
d(d-1)\[\dot{H}+\frac{1}{2}(d+1)H^2\]\, ,\label{UV1}\\
&&\ddot{V}+\mu^2V+dH\dot{V}-2(d+1) H^2V=
-(d-1)\dot{H}\, ,\label{UV2}
\ees
which provides the generalization of the Friedmann equation to non-local massive 
gravity.\footnote{Note that, for  $\mu^2=0$,  \eq{UV1} can be rewritten as
$a^{-d}\pa_0( a^d\dot{U})=d(d-1)a^{-(d+1)/2}\pa_0( a^{(d+1)/2}H)$,
which integrates to
$\dot {U}(t)=d(d-1) a^{2d}(t)H(t)-\frac{d(d-1)^2}{2} a^d(t)
\int_0^t dt' a^d(t') H^2(t')$.}
Observe that $U$ and $V$ enter as new propagating degrees of freedom, corresponding to the two dynamical degrees of freedom in the scalar sector, and therefore one must also impose appropriate initial conditions on them. Further initial data are required for the inversion of the operator $(\pa_0^2+dH\pa_0-dH^2)$.
A detailed study of the solutions of these equations will be presented in subsequent work.

\vspace*{5mm}

\noindent {\bf Acknowledgments}.
We thank   Claudia de~Rham, Alberto Nicolis, Antonio Riotto, Andrew Tolley and Gabriele Veneziano for very useful discussions and comments on the manuscript.
Our work is supported by the Fonds National Suisse.

\appendix

\section{Properties of $\iBox$ and $\iBox_g$}\label{app:iBox}

In this appendix we recall some elementary facts on the inverse d'Alembertian in flat and in  curved space. We begin with the flat-space d'Alembertian, that we denote simply as $\Box$, while we reserve the notation $\Box_g$ for the d'Alembertian in the metric $\gmn$. The general solution of an equation of the form
$\Box\varphi=j$ is
\be\label{solphiG}
\varphi(x)\equiv  (\Box^{-1}j)(x)=\varphi_{\rm hom}(x)+\int d^4x' G(x-x') j(x')
\, ,
\ee
where  $\varphi_{\rm hom}(x)$ is a solution of the homogeneous equation and $G(x-x')$ is a Green's function of the d'Alembertian operator,
\be\label{defGreen}
\Box_xG(x-x')=\d^{(4)}(x-x')\, .
\ee
Observe that, with the signature $(-,+,+,+)$, the propagator $D(x)$
of the quantum theory is defined by $\Box_xD(x-x')=i\d^{(4)}(x-x')$, so $D(x)=iG(x)$.
Solutions corresponding to different Green's functions differ by a solution of the homogeneous equation. In all the formal manipulations involving $\iBox$ we will set to zero the homogeneous solution. In this way the kernel of the $\Box$ operator becomes trivial, its inversion is well defined and we can perform a number of formal operations, such as integrating $\iBox$ by parts, see below.
The choice of the Green's function is determined  by  the physics of the problem.
At the classical level, causality  requires the use of the
retarded Green's function
\be\label{Gret}
G_{\rm ret}(x;x')=-\frac{1}{4\pi}\, \frac{1}{|\vx-\vx'|}
\d \( t-t' -|\vx-\vx'|\)\, .
\ee
The advanced Green's function is instead
\be\label{Gadv}
G_{\rm adv}(x;x')=-\frac{1}{4\pi}\, \frac{1}{|\vx-\vx'|}
\d \( t-t' +|\vx-\vx'|\)\, .
\ee
In flat space, all Green's functions are actually a function of $x-x'$ only. However, of course, $G_{\rm ret}(x;x')$ and $G_{\rm adv}(x;x')$ are not  symmetric in $x,x'$. Rather exchanging $x$ with $x'$,
$G_{\rm ret}(x;x')$ becomes $G_{\rm adv}(x;x')$, and viceversa.  A Green's function invariant under $x\leftrightarrow x'$ is obtained  taking the symmetric combinations
\be
G_{+}(x;x')=\frac{1}{2}\[G_{\rm ret}(x;x')+G_{\rm adv}(x;x')\]
=-P\int \frac{d^4p}{(2\pi)^4}\, \frac{1}{p^2}\, e^{-ip(x-x')}\, ,
\ee
where $P$ denotes the principal part. Another Green's functions invariant under $x\leftrightarrow x'$ is the Feynman Greens' function
\be
G_{F}(x;x')
=-\int \frac{d^4p}{(2\pi)^4}\, \frac{1}{p^2-i\eps}\, e^{-ip(x-x')}\, .
\ee
The two differ by an imaginary term, according to the relation $1/(y\pm i\eps)=
P(1/y)\mp i\pi \d(y)$.
Observe that the operator $\Box^{-1}$ is self-adjoint only if it is defined using a symmetric Green's function, i.e. $G_+$ or $G_F$,  since in this case  for any two differentiable and square integrable functions $A(x)$ and $B(x)$ we have (using  for definiteness $G_+$)
\be
\int d^4x\, A(x)({\Box}^{-1}_{+} B)(x)=\int d^4x d^4x' \, A(x)G_+(x;x')B(x')=\int d^4x\, ({\Box}_{+}^{-1}A)(x) B(x)\, .\label{intpartokflat}
\ee
or, in other words, $\Box_{+}^{-1}$ can be integrated by parts.
Observe also that in flat space, for a generic Green's function, $\pam$ commutes with $\iBox$. This is a consequence of the fact that in flat space $G(x;x')=G(x-x')$. Thus
\bees
\pam (\iBox f)(x)&=&\int d^4x'\, \[\frac{\pa}{\pa x^{\mu}}G(x-x')\]f(x')\nn\\
&=& - \int d^4x'\,\[\frac{\pa}{\pa {x'}^{\mu}}G(x-x')\]f(x')
= + \int d^4x'\, G(x-x')\frac{\pa}{\pa {x'}^{\mu}}f(x')\nn\\
&=&\iBox (\pam f)(x)\, ,\label{pacommiBox}
\ees
where in the second line we integrated $\pa/\pa {x'}^{\mu}$ by parts.

We next consider the inverse d'Alembertian in curved space.
On a scalar function $f$ the inverse of
$\Box_g$ is defined by
\be\label{defboxinve}
(\Box_g^{-1}f)(x)=\int dx' \sqrt{-g(x')}\,\, G_g(x;x') f(x')\, ,
\ee
where 
\be\label{BoxgG}
(\Box_g)_x G_g(x;x')=\frac{1}{\sqrt{-g(x)}}\d(x-x')\, ,
\ee
and
\be\label{Boxscalar}
\Box_g=\frac{1}{\sqrt{-g}}\pam (\sqrt{-g}\, \gMN\pan)
\ee
is the d'Alembertian on a scalar function.
The notation $(\Box_g)_x$ indicates that the derivatives are with respect to $\xm$.  The factor $1/\sqrt{-g(x)}$ on the right-hand side 
of \eq{BoxgG}
is chosen because  under coordinate transformation $\d(x-x')/\sqrt{-g(x)}$ is a scalar (rather than a scalar density), as it is clear from the fact that its integral over $dx\sqrt{-g}$ is equal to one. Thus, with this definition even $G_g(x;x')$ and hence $\Box_g^{-1}f$ are scalar under coordinate transformations.

Observe that in a generic space-time the Green's function is no longer a function of the difference $x-x'$. However, $G_F(x;x')$ and $G_+(x;x')$ are still symmetric under the exchange of $x$ and $x'$. The operator $\Box_g^{-1}$ defined using  $G_{g,+}(x;x')$ or
$G_{g,\rm F}(x;x')$ is therefore self-adjoint and we can integrate it by parts,
since in this case
\bees
&&\int d^4x\,\sqrt{-g(x)}\, A(x)({\Box}^{-1}_g B)(x)=\int d^4x \sqrt{-g(x)}
d^4x' \sqrt{-g(x')}\, A(x)G_{g}(x;x')B(x')\nn\\
&&=\int d^4x \sqrt{-g(x)}d^4x' \sqrt{-g(x')}
\, A(x)G_{g}(x';x)B(x')\nn\\
&&=
\int d^4x \sqrt{-g(x)}\, B(x) \int d^4x'\sqrt{-g(x')}\, G_{g}(x;x') A(x')\nn\\
&&=\int d^4x\sqrt{-g(x)}\, ({\Box}^{-1}_gA)(x) B(x)\, .\label{intpartokcurved}
\ees
Of course the $\Box_g$ operator depends on whether it acts on a scalar,  a scalar density, a four-vector, a tensor, etc, and the same is true for $\Box_g^{-1}$.
Observe that,
 since $\gmn$ commutes with $\n_{\mu}$, it commutes also with $\Box$. Therefore, for any tensor $\Tmn$,
\be
\gMN\Tmn=\Box(\gMN\iBox\Tmn)
\ee
Applying $\iBox$ to both sides,
\be\label{App:aaa1}
\iBox(\gMN\Tmn)=\gMN\iBox\Tmn\, .
\ee
Now $\gMN\Tmn$ is a scalar, so
\be
\iBox(\gMN\Tmn)=\int dx' \sqrt{-g(x') }\,G(x;x')\gMN(x')\Tmn(x')\, ,
\ee
where $G(x;x')$ is a Green's function of the  $\Box_g$ operator acting on scalars.
Thus, the definition of $\iBox$ on a tensor $\Tmn$ is such that
\be\label{App:aaa2}
\gMN(x)(\iBox\Tmn)(x) =\int dx' \sqrt{-g(x') }\,G(x;x')\gMN(x')\Tmn(x')\, .
\ee
The explicit form of $\Box_g^{-1}$ on a scalar density (such as $\sqrt{-g}R$) can be obtained similarly, observing that  $\n_{\mu}(\sqrt{-g})=0$. Thus, $\sqrt{-g}$ commutes with $\Box_g=\n_{\mu}\n^{\mu}$, and this implies that it also commutes with
$\Box^{-1}_g$.\footnote{The proof is obtained writing $\n_{\mu}(\sqrt{-g}\,\Box^{-1}_g R)=\sqrt{-g}\,\n_{\mu}(\Box^{-1}_g R)$.
Applying again $\n^{\mu}$, $\Box_g
(\sqrt{-g}\,\Box^{-1}_g R)=\sqrt{-g}\,\Box_g\Box^{-1}_g R=\sqrt{-g}R$. 
With the definition of $\Box_g$ that sets to zero the solution of the homegeneous equation we also have $\Box_g^{-1}\Box_g=1$, so we get
$\sqrt{-g}\, \Box^{-1}_g R=\Box^{-1}_g (\sqrt{-g}\,R)$. }
This means that the  definition of $\Box^{-1}_g$ on a scalar density, such as $\sqrt{-g}R$, is
\bees
[\Box^{-1}_g(\sqrt{-g}R)](x)&\equiv&\sqrt{-g(x)} (\Box^{-1}_gR)(x)\nn\\
&=&\sqrt{-g(x)}\int dx' \sqrt{-g(x')}\,\, G_g(x;x') R(x')\, .
\ees
Thus, $\Box^{-1}_g$ applied to a scalar density gives back a scalar density. We
can therefore write equivalently $\sqrt{-g}\Box^{-1}_gR$ or $\Box^{-1}_g(\sqrt{-g}R)$, taking however in account the different definitions of $\Box^{-1}_g$ on scalars and on scalar densities.

\section{Non-local field redefinitions and  propagating degrees of freedom}\label{sect:dyn}

Non-local field theories are certainly less familiar than the usual local field theories, and when  manipulating them one must be aware of some subtleties.
In particular, blind manipulations might lead one to believe that the theory has more propagating degrees of freedom than it actually has, and might even lead one to believe that the theory has ghosts when in fact it is perfectly healthy. The issue is interesting in itself, and is important for understanding the degrees of freedom in the non-local formulations of massive gravity, so we discuss it here in some detail.

As a first trivial example consider a scalar field $\phi$ which  satisfies a non-dynamical equation of motion such as a Poisson equation  $\n^2\phi=\rho$. If we  define a new field $\tilde{\phi}$ from  $\tilde{\phi}=\iBox\phi$,  the equation of motion  can be written as  $\Box\tilde{\phi}=\n^{-2}\rho\equiv\tilde{\rho}$, and now $\tilde{\phi}$ looks like a dynamical field. However, we certainly cannot transform a non-dynamical   degree of freedom into a dynamical one in this manner. A way to see where the procedure goes wrong is to realize that
assigning initial conditions on a given time slice to $\phi$ does not provide initial conditions on $\tilde{\phi}$. To get 
$\tilde{\phi}$ on any single time slice, we need to know $\phi$ everywhere not only in space but even in time. Alternatively, we can observe that  for $\rho=0$ we have $\phi=0$, which means that the we must set also $\tilde{\phi}=0$ in order not to introduce spurious degrees of freedom. In other words, among  the solutions of the equation $\Box\tilde{\phi}=0$, we must discard all the plane-wave solutions, and only retain $\tilde{\phi}=0$. 

An example of this sort appears even in linearized massless GR, when we decompose the metric perturbation in terms of 
quantities which are transverse or longitudinal with respect to the Lorentz group. The decomposition reads
\be\label{decomphmn}
\hmn =\hmn^{\rm TT}+\frac{1}{2}(\pam \eps_{\nu}^{\rm T}+\pan \eps_{\mu}^{\rm T}) +
\pam\pan\a+\frac{1}{d}\emn s\, ,
\ee
where $\hmn^{\rm TT}$ is transverse and traceless with respect to the Lorentz indices,
\be\label{prohjh2}
\paM \hmn^{\rm TT}=0\, ,\qquad \eMN \hmn^{\rm TT}=0\, ,
\ee
and 
$\paM \eps_{\mu}^{\rm T}=0$. The factor $1/d$ in front of $s$ is an unconventional normalization that will be useful later.
Thus, in $d=3$, $\hmn^{\rm TT}$ carries five degrees of freedom, $\eps_{\mu}^{\rm T}$ three, and two  scalar degrees of freedom are carried by $\a$ and $s$. 
Observe that $\eps_{\mu}^{\rm T}$ and $\a$ come from the decomposition  of a generic four-vector
$\eps_{\mu}=\eps_{\mu}^{\rm T}+\pam\a$.
It is  straightforward to invert \eq{decomphmn} and express $\a,s,\eps_{\mu}^{\rm T}$ and $\hmn^{\rm TT}$ in terms of
$\hmn$, but  the inversion involves the non-local operator $\iBox$.
The explicit expression of $s$ and $\a$ in terms of $\hmn$ can be found taking the trace of 
\eq{decomphmn}, which gives
 $h=[(d+1)/d]s+\Box\a$, and contracting \eq{decomphmn} with $\paM\paN$, which gives
 $\paM\paN\hmn=\Box [(s/d)+\Box\a]$. Combining these equations we get 
 \bees
 s&=&\(\eMN-\frac{1}{\Box}\paM\paN\)\hmn\, ,\label{defshmn}\\
 \a&=&-\frac{1}{d}\, \frac{1}{\Box}\(\eMN-\frac{d+1}{\Box}\paM\paN\)\hmn\, .
 \ees
We can now extract $\eps_{\mu}^{\rm T}$ by applying $\paM$ to \eq{decomphmn} and using the above expressions for $\a$ and $s$. This gives
\be
\eps_{\mu}^T=\frac{2}{\Box}\(\d_{\mu}^{\rho}-\frac{\pam\paR}{\Box}\)\paS\hrs\, .
\ee
Finally, substituting these expressions into \eq{decomphmn} we get
\bees
\hmn^{\rm TT}&=&\hmn -\frac{1}{d}\(\emn -\frac{\pam\pan}{\Box}\)h
-\frac{1}{\Box}(\pam\paR\hnr+\pan\paR\hmr)+
\frac{1}{d}\,\emn \frac{1}{\Box}\paR\paS\hrs \nn\\
&&+\frac{d-1}{d}\frac{1}{\Box^2}\pam\pan\paR\paS\hrs\, .\label{defhath}
\ees
Observe also that $\hmn^{\rm TT}$ and $s$ are invariant under linearized diffeomorphisms, while the four-vector $\eps_{\mu}=\eps_{\mu}^{\rm T}+\pam\a$ transforms as
$\eps_{\mu}\ra \eps_{\mu}-\xi_{\mu}$. Thus we can choose the gauge so that $\eps_{\mu}=0$, and this leaves no residual gauge symmetry.

The crucial point is that this inversion involves $\iBox$ and is therefore non-local both in space and time,\footnote{This should be contrasted with the usual $(3+1)$ decomposition of the metric, which involves only the inversion of the Laplacian $\n^2$
(see e.g. \cite{Flanagan:2005yc,Jaccard:2012ut}), and is therefore non-local in space but local in time.}  and a blind use of these variables can lead to some apparent paradox.
 Indeed,
substituting \eq{decomphmn} in the quadratic Einstein-Hilbert action, $\eps_{\mu}$ drops because of the invariance under linearized diffeomorphisms, and one finds
\be\label{SEH2nl}
S_{\rm EH}^{(2)}=\frac{1}{2}\int d^{d+1}x \, h_{\mu\nu} {\cal E}^{\mu\nu,\rho\sigma} h_{\rho\sigma}=\frac{1}{2}\int d^{d+1}x \,\[\hmn^{\rm TT}\Box (h^{\mu\nu })^{\rm TT}
-\frac{d-1}{d}\, s\Box s\]\, .
\ee
Performing the same decomposition in the energy-momentum tensor, the interaction term 
can be written as\footnote{Writing 
$\Tmn =\Tmn^{\rm TT}+(1/2)(\pam S_{\nu}^{\rm T}+\pan S_{\mu}^{\rm T}) +
\pam\pan \Sigma +\emn S$,
energy-momentum conservation implies 
$(1/2)\Box S_{\nu}^{\rm T} +
\pan(\Box \Sigma + S)=0$.
The transverse and longitudinal parts of this expression must vanish separately. For a localized source, from $\Box S_{\nu}^{\rm T}=0$  it follows that
$S_{\nu}^{\rm T}=0$. Eliminating $S$ from $S=-\Box\Sigma$ and expressing $\Sigma $ in terms of $T$ using $T=\Box \Sigma+(d+1)S=-d\Box\Sigma$ it follows that
$\Tmn =\Tmn^{\rm TT}+(1/d)( \emn-\iBox\pam\pan)T$, which gives \eq{Sint2}.
}
\be\label{Sint2}
S_{\rm int}=\frac{\kappa}{2}\int d^{d+1}x\, \hmn\TMN 
=\frac{\kappa}{2}\int d^{d+1}x\,\[ \hmn^{\rm TT}(\TMN)^{\rm TT}+\frac{1}{d}sT\]
\, ,
\ee
so the equations of motion derived from $S_{\rm EH}^{(2)}+S_{\rm int}$ are
\be
\Box\hmn^{\rm TT}=-\frac{\kappa}{2}\Tmn^{\rm TT}\, ,\qquad
\Box s=+\frac{\kappa}{2(d-1)}T\, .
\ee
Thus, using these variables one might be induced to conclude that linearized GR has 
six radiative degrees of freedom, because both
$\hmn^{\rm TT}$ and $s$ are governed by a KG equation. Observe furthermore that these degrees of freedom are gauge invariant, so they cannot be  gauged away. Furthermore, for all $d>1$ the scalar $s$ has the ``wrong" sign of the kinetic term in the action, so one might be induced to conclude that it is a ghost. 

Of course these conclusions are wrong, and linearized GR is a ghost-free theory with only two radiative degrees of freedom, corresponding to the $\pm 2$ helicities of the graviton. The loophole in the above argument is exactly the same as in the trivial example presented at the beginning of this section, where a non-dynamical field $\phi$ was transformed into an apparently  dynamical field $\tilde{\phi}$ through the redefinition $\tilde{\phi}=\iBox\phi$. Indeed, expressing $s$ 
in terms of the variables entering the $(3+1)$ decomposition, we find (specializing the above results to $d=3$)
$s=6\Phi- 2\iBox\n^2(\Phi+\Psi)$,
where $\Phi$ and $\Psi$ are the scalar Bardeen's variable defined in flat space (see \cite{Jaccard:2012ut}). Since $\Phi$ and $\Psi$ are non-radiative,  $s$ is non-radiative too.  The fact that it is obtained applying the $\iBox$ operator to the non-radiative field $\n^2(\Phi+\Psi)$ gives to its equation  of motion the appearance  of a dynamical equation, but nevertheless $s$ does not represent a dynamical degree of freedom of the theory. Again, this is reflected in the fact that giving initial conditions on a given time slice for the metric does not provide the initial conditions on $s$.

Another example of the apparent puzzles that can arise from non-local field redefinitions
is obtained diagonalizing the action (\ref{SmassiveGNonLoc}). 
Following \cite{Hinterbichler:2011tt}, the quadratic term that mixes $N$ and $\hmn$ can be removed   defining\footnote{In generic $d$ spatial dimensions, all  equations written in Sect.~\ref{sect:NLFP}
simply go through with the trivial replacement $d^4x\ra d^{d+1}x$. In contrast, the space-time dimension enters in the diagonalization, and in $d$ spatial dimensions the factor $\sqrt{6}$ in the equations below must be replaced by 
$\sqrt{d(d-1)}$.} 
\bees
h'_{\mu\nu}&=&\hmn-\emn\frac{m^2}{\Box-m^2}N\, ,\label{diagonal1}\\
N'&=&\sqrt{6}\, \frac{m^2}{\Box-m^2}N\, .\label{diagonal2}
\ees
The action (\ref{SmassiveGNonLoc}) then becomes~\cite{Hinterbichler:2011tt}
\bees\label{SmassiveGNonLochatN}
S_{\rm FP}+S_{\rm int}&=& \int d^4x\, \[
\frac{1}{2}h'_{\mu\nu} \(1-\frac{m^2}{\Box}\) {\cal E}^{\mu\nu,\rho\sigma}
h'_{\rho\sigma}
+\frac{1}{2}N'(\Box-m^2)N'\]\nn\\
&&+\frac{\kappa}{2}  \int d^4x\, \(h'_{\mu\nu}\TMN
+\frac{1}{\sqrt{6}}N'T\)
\, ,
\ees
where $T=\eMN\Tmn$. At first sight something very strange happened here, since the term
$h'_{\mu\nu} (1-m^2/\Box) {\cal E}^{\mu\nu,\rho\sigma}h'_{\rho\sigma}$ has the same functional form as the term 
$h_{\mu\nu} (1-m^2/\Box) {\cal E}^{\mu\nu,\rho\sigma}h_{\rho\sigma}$, so one would think that the two  describe the same number of dynamical degrees of freedom. However, the field $N$, which 
in \eq{SmassiveGNonLoc} entered as a Lagrange multiplier and removed a scalar degree of freedom from $h_{\mu\nu} (1-m^2/\Box) {\cal E}^{\mu\nu,\rho\sigma}h_{\rho\sigma}$, has now been traded for a field $N'$ which looks fully dynamical. Thus, we have apparently lost a scalar constraint, and furthermore we have gained a dynamical scalar field, so the number of scalar degrees of freedom  apparently increased by two. 

Again, the solution to this apparent puzzle is that the correct counting of radiative degrees of freedom can only be done using the original variables $\hmn$ and $N$.\footnote{Observe that $N$ is a combination of $h$ and of $\Box\a =-\pam\AMU$, so it does not involve non-local operators.} For instance, the field $N'$ is a fake dynamical field, just as the field $\tilde{\phi}$ discussed at the beginning of this section. Indeed, $N$ is determined algebraically by \eq{NTlin}, 
$N=c T$, with $c$ a constant, so  \eq{diagonal2} gives $(\Box-m^2)N'=c' \, T$. However,
the initial conditions on $\{\hmn,N\}$ do not fix the initial conditions on 
$\{\hmn',N'\}$. Rather, $\hmn'$ and $N'$ at a given time slice can only be determined if we know $\hmn$ and $N$ at all times, i.e. if we have already solved the equations of motion. 
Of course, nothing forbids one from considering the theory (\ref{SmassiveGNonLochatN}) for its own sake, and solve its  equations of motion assigning initial conditions on $\{\hmn',N'\}$. However, in this way we define a different  theory, which has nothing to do with FP massive gravity, even as far as the number of dynamical degrees of freedom is concerned.

One should also be careful in the use of an action such as (\ref{SEH2nl}) and of its non-linear extension, when computing the S-matrix elements. Indeed, while standard theorems assure the invariance of the $S$-matrix under local field redefinitions, its invariance under non-local field redefinitions is in general not assured.

\section{The action of the massive theory as a local functional of non-local fields}
\label{app:nonlocfields}

We have seen in the previous appendix that non-local transformations of the field must be used with care, particularly when one wishes to study  what are the dynamical and  non-dynamical degrees of freedom of the theory. Having understood this point, it is however still interesting to observe that there exist non-local transformations of the fields that bring the non-local actions that we have discussed into simple and elegant local forms, which can be useful for obtaining a further understanding of the structure of these theories. In particular we will see explicitly how, for $m\ra 0$, the ghost of the massive theory smoothly reduces to a non-radiative degree of freedom of GR.
We start  again from  electrodynamics, where the construction is simpler. 

\subsection{Non-local variables in  electrodynamics}\label{sect:hath}

In electrodynamics  the gauge field $\Am$ can be separated into a transverse and a longitudinal part,
\be\label{separAMU}
\Am=\Am^{\rm T}+\pam\a\, ,
\ee
where 
\begin{equation}  \label{pamhatAM}
\paM \Am^{\rm T} = 0 \, .
\end{equation}
To invert \eq{separAMU} we take its divergence, which gives
$\paM \Am=\Box\a$, so that
\be
\a=\iBox \paM \Am\, .
\ee
Substituting this into $\Am^{\rm T}=\Am-\pam\a$ we get
\be\label{defAhat}
\Am^{\rm T}=\Am-\frac{1}{\Box}\pam\paN\An=
P_{\mu}^{\nu}A_{\nu}\, ,
\ee 
where we introduced the non-local operator
\begin{equation}  \label{defPMunu}
P_{\mu}^{\nu} \equiv \delta_{\mu}^{\nu} - \frac{\partial_{\mu} \partial^{\nu}}{\Box}\, .
\end{equation}
Observe that $P_{\mu}^{\rho}P_{\rho}^{\nu}=P_{\mu}^{\nu}$.
Furthermore, applying $P_{\mu}^{\nu}$ to a pure gauge configuration we get zero,
\begin{equation}  \label{Ppamth}
P_{\mu}^{\nu} \partial_{\nu} \theta = 0 \, .
\end{equation}
Since $P_{\mu}^{\nu}$ is linear, this implies that $\Am^{\rm T}$ is gauge-invariant,
\be
\Am^{\rm T}\ra P_{\mu}^{\nu}( A_{\nu}-\pan\theta)=\Am^{\rm T}\, .
\ee
Thus, under a gauge transformations $\Am\ra\Am-\pam\theta$, we have 
$\a\ra\a-\theta$ and  $\Am^{\rm T}\ra \Am^{\rm T}$.
Thus $P_{\mu}^{\nu}$ is a projector that associates to a gauge orbit (of which $A_{\mu}$ is a representative)  a gauge-invariant vector
field $\Am^{\rm T}$  that satisfies the Lorentz condition. 
Observe that, because of \eq{pamhatAM},  $\Am^{\rm T}$ describes three degrees of freedom. In the case of massive electrodynamics these are the three spin states of a massive photon. Thus, $\Am^{\rm T}$ provides a {\em gauge-invariant} description of the  three physical degrees of freedom of massive electrodynamics. We see here again the interplay between gauge-invariance and locality in the massive gauge theory. If we insist on manifest locality we must use the gauge-field $\AMU$, which is not gauge-invariant, and we cannot construct with it a local gauge-invariant mass term. In contrast, if we give up manifest locality, we have at our disposal a field $\Am^{\rm T}$ which is gauge invariant. Using this field
 it is straightforward to write an action with  a mass term that does not spoil  gauge-invariance,
\be\label{eq:massMaxw}
S_{\rm gauge-inv} = \int d^4 x \left( -\frac{1}{4} F_{\mu\nu} F^{\mu\nu} - \frac{1}{2} m_{\g}^2 \Am^{\rm T} A^{T\mu} \right) -j^{\mu}A_{\mu} \, .
\ee
We could have also replaced $A_{\mu} \to \Am^{\rm T}$ in the kinetic term.
However, we see from \eq{separAMU} that
\be
F_{\mu\nu}=\pam A_{\nu}-\pan A_{\mu}=
\pam A_{\nu}^T-\pan A_{\mu}^T
\, .
\ee
Similarly, upon integration by parts, for a conserved current we can write equivalently $j^{\mu}A_{\mu}$
or $j^{\mu}\Am^{\rm T}$.
We now insert into (\ref{eq:massMaxw})
the non-local expression of $\Am^{\rm T}$ in terms of $\Am$ given in  (\ref{defAhat}). Performing some integration by parts  and using the identity
\be\label{identi}
\Fmn\frac{1}{\Box}\FMN= 2(\pam\An)\frac{1}{\Box}(\paM\AN-\paN\AMU)\nn\\
=-2\An\AN -2(\pam\AMU)\frac{1}{\Box}(\pan\AN)\, ,
\ee
(in which again the second equality has been obtained integrating by parts)
we find that
\be\label{hatAhatA}
\Am^{\rm T} A^{T\mu}=-\frac{1}{2}\Fmn\frac{1}{\Box}\FMN\, ,
\ee
and therefore
\be
S_{\rm gauge-inv}=-\frac{1}{4}\int d^4x \, \Fmn \(1-\frac{m_{\g}^2}{\Box}\)\FMN
\, .
\ee
Thus, the non-local action (\ref{Lnonloc}) is equivalent to (\ref{eq:massMaxw}). In the former action the non-locality is explicitly displayed. In the latter, it is hidden in the non-local relation between $\Am^{\rm T} $ and $\Am$.
We stress again  that this non-locality has no physical consequences, since it only affects pure gauge modes, and can be gauged away giving up explicit gauge invariance by fixing the Lorentz gauge $\paM\Am=0$, since in this gauge $\Am^T$ reduces to the local field $\Am$, as we see from
\eq{defAhat}. This illustrates again  the interplay between gauge invariance  and locality in massive electrodynamics.

It is also interesting to note that the projection onto the transverse part allows one to define a scalar product in the physical configuration space, i.e. in the space of gauge orbits. Indeed, the following bilinear functional is independent of the orbit representative
\begin{equation}  \label{eq:physconfmetric}
\langle A,B \rangle \equiv \frac{1}{2}\int d^4x \, \Am^{\rm T} {B}^{T\mu} \, ,
\end{equation}
since we have seen that $\Am^{\rm T}$ and $ {B}^{T\mu}$ are gauge-invariant.
Using \eq {hatAhatA} we see that the Proca action can be written as the ``expectation value" of the Klein-Gordon operator with respect to that scalar product
\begin{equation}  \label{eq:massiveMax}
S_{\rm gauge-inv} = \langle A, (\Box - m^2) A \rangle \, .
\end{equation}
In terms of this scalar product the inclusion of a mass term is therefore trivial. 

\subsection{Non-local variables in linearized gravity}\label{app:NLgra}

We now generalize to the spin-2 case the construction of non-local variables discussed above. 
We begin by introducing a projector $P_{\mu\nu}^{\rho\sigma}$ which is
just the symmetrization of the square of the projector $P_{\mu}^{\nu}$ defined in
\eq{defPMunu},
\bees 
P_{\mu\nu}^{\rho\sigma} &\equiv& 
\frac{1}{2}
\( P_{\mu}^{\rho}P_{\nu}^{\sigma} +P_{\nu}^{\rho}P_{\mu}^{\sigma} \)\label{defPP}\\
&=& 
\frac{1}{2}(\delta_{\mu}^{\rho} \delta_{\nu}^{\sigma}+\delta_{\nu}^{\rho} \delta_{\mu}^{\sigma})
 - \frac{1}{2\Box} (\delta_{\mu}^{\rho} \partial_{\nu} \partial^{\sigma}
+ \delta_{\nu}^{\sigma} \partial_{\mu} \partial^{\rho}+
\delta_{\nu}^{\rho} \partial_{\mu} \partial^{\sigma}
+ \delta_{\mu}^{\sigma} \partial_{\nu} \partial^{\rho})
 + \frac{1}{\Box^2} \partial_{\mu} \partial_{\nu} \partial^{\rho} \partial^{\sigma} \, .\nn
\ees
We can then define a projected field 
\be\label{defhhmn}
\hhmn \equiv P_{\mu\nu}^{\rho\sigma}\hrs\, .
\ee
Observe that $\hat{h}_{\mu\nu}$ can be obtained starting from $\hmn$ and performing a gauge transformation 
$\hmn\ra \hmn -(\pam\xin+\pan\xim)$ with 
\be
\xin=\frac{1}{\Box}\paR\hnr
-\frac{1}{2\Box^2}\pan (\paR\paS\hrs)\, .
\ee
Similarly to the spin-1 case, the projector  gives zero on pure-gauge configurations,
\be
P_{\mu\nu}^{\rho\sigma} \left( \partial_{\rho} \xi_{\sigma} + \partial_{\sigma} \xi_{\rho} \right) 
= 0 \, ,\label{prohjh1}
\ee
and therefore $\hat{h}_{\mu\nu}$ is gauge-invariant under linearized diffeomorphisms,
\be
\hhmn\ra P_{\mu\nu}^{\rho\sigma}\[\hrs -(\partial_{\rho} \xi_{\sigma} + \partial_{\sigma} \xi_{\rho} )\]=\hhmn\, .
\ee
Thus, in full analogy with the case of electrodynamics, $P_{\mu\nu}^{\rho\sigma}$ sends $\hmn$ into a gauge-invariant field $\hhmn$, that satisfies the transversality condition $\paM\hhmn=0$, and which is a gauge-invariant representative of the gauge orbit to which $\hmn$ belongs. 
The field $\hhmn$ is transverse but not traceless. Defining $\hat{h}=\eMN\hhmn$, we have the identity
\be\label{defHmunu}
\hmn^{\rm TT}=\hhmn-\frac{1}{d}\(\emn-\frac{1}{\Box}\pam\pan\)\hat{h}\, .
\ee
Contracting with $\eMN$ or with
$\paM$, and using $\paM\hhmn$, we see that the right-hand side of 
\eq{defHmunu} is indeed transverse and traceless so it is clear that it must be equal to $\hmn^{\rm TT}$. This   can indeed be immediately checked using the explicit expressions of 
$\hhmn$ given in \eqs{defPP}{defhhmn} and comparing with the explicit expression of 
$\hmn^{\rm TT}$ given in \eq{decomphmn}.
Since $\hhmn$ is invariant under linearized diffeomorphisms, \eq{defHmunu}   nicely shows that
$\hmn^{\rm TT}$ is also invariant under diffeomorphisms. 

From \eq{defshmn} we see that also $s$ is invariant under linearized diffeomorphisms. In contrast, as it is clear from \eq{decomphmn}, the vector $\eps_{\mu}=\eps_{\mu}^{\rm T}+\pam\a$ transforms as $\eps_{\mu}\ra \eps_{\mu}-\xi_{\mu}$, and can be set to zero by a gauge transformation. Observe that choosing the gauge so that $\eps_{\mu}=0$ leaves no residual gauge freedom.
Thus, $\eps_{\mu}$ describes the four pure-gauge modes, while the five degrees of freedom of $\hmn^{\rm TT}$, together with the scalar  $s$, describe six physical degrees of freedom of the gravitational field. As we learned in App.~\ref{sect:dyn}, these variables are not appropriate for identifying which degrees of freedom are radiative and which are not. However, we already saw in sect.~\ref{sect:dof} that, in the theory defined by the action $S_{\rm non-loc}$, all the six gauge-invariant degrees of freedom are radiative. 

Just as in the case of massive electrodynamics,   it is straightforward to write a gauge-invariant action for linearized massive gravity using the gauge-invariant variable $\hhmn$.
We can in fact construct the gauge-invariant action
\be
S_{\rm gauge-inv}=\frac{1}{2}\int d^{d+1}x \left[{h}_{\mu\nu} {\cal E}^{\mu\nu,\rho\sigma} {h}_{\rho\sigma} - m^2 \left(  \hat{h}_{\mu\nu}\hat{h}^{\mu\nu} - \hat{h}^2 \right) \right] \, ,
\ee
with a FP mass term constructed using $\hat{h}_{\mu\nu}$ 
and 
$\hat{h}\equiv \eMN\hat{h}_{\mu\nu}$. Observe that
\be\label{hEhhEh}
{h}_{\mu\nu} {\cal E}^{\mu\nu,\rho\sigma} {h}_{\rho\sigma}=
\hat{h}_{\mu\nu} {\cal E}^{\mu\nu,\rho\sigma} \hat{h}_{\rho\sigma}\, ,
\ee 
since this term is gauge-invariant and $\hmn$ and $\hhmn$ are related by a gauge 
transformation.
From the explicit  expression of $\hat{h}_{\mu\nu}$ in terms of $\hmn$ given by \eq{defhath} we find that 
\be
\int d^{d+1}x\, \,\(\hat{h}_{\mu\nu}\hat{h}^{\mu\nu} - \hat{h}^2 \)=
\int d^{d+1}x\, \,\(\hmn \, \frac{1}{\Box}
{\cal E}^{\mu\nu,\rho\sigma}{h}_{\rho\sigma}\)
\, ,
\ee
and therefore
$S_{\rm gauge-inv}$ is the same as the non-local action $S_{\rm non-loc}$ given in \eq{Snon-loc}. We therefore have the identities
\bees\label{Snon-loc2}
S_{\rm non-loc}&\equiv& \int d^{d+1}x\, \,  \frac{1}{2}
{h}_{\mu\nu} \(1-\frac{m^2}{\Box}\) {\cal E}^{\mu\nu,\rho\sigma}
{h}_{\rho\sigma}\nn\\
&=&\int d^{d+1}x \left[\frac{1}{2}\hat{h}_{\mu\nu} {\cal E}^{\mu\nu,\rho\sigma} \hat{h}_{\rho\sigma} - \frac{m^2}{2} \left(  \hat{h}_{\mu\nu}\hat{h}^{\mu\nu} - \hat{h}^2 \right) \right] \nn\\
&=&\int d^{d+1}x\, \[  \frac{1}{2}\hat{h}_{\mu\nu}(\Box-m^2)\hat{h}^{\mu\nu}
- \frac{1}{2}\hat{h}(\Box-m^2)\hat{h}\]\, ,
\ees
where in the last line  we used \eq{prohjh2} to simplify $\hat{h}_{\mu\nu} {\cal E}^{\mu\nu,\rho\sigma} \hat{h}_{\rho\sigma}$. Thus, $S_{\rm non-loc}$ is a {\em local\,} functional of $\hhmn$, and the non-locality of $S_{\rm non-loc}$ as a functional of the metric perturbation $\hmn$ is now hidden in the non-local relation between $\hhmn$ and $\hmn$.

In the action (\ref{Snon-loc2}) the terms $\hat{h}_{\mu\nu}(\Box-m^2)\hat{h}^{\mu\nu}$ and $\hat{h}(\Box-m^2)\hat{h}$ are not independent, since $\hat{h}=\eMN\hat{h}_{\mu\nu}$. We can however decouple them using \eq{defHmunu} to write
\be
\hhmn=\hmn^{\rm TT}+\frac{1}{d}\(\emn-\frac{1}{\Box}\pam\pan\)\hat{h}\, .
\ee
and we can use $\hmn^{\rm TT}$ and $\hat{h}$ as independent fields.
Then the action (\ref {SFPhath}) becomes
\be
S_{\rm non-loc}=  \frac{1}{2}\int d^{d+1}x\, \[ h^{\rm TT}_{\mu\nu}(\Box-m^2)h^{{\rm TT}\mu\nu}
- \frac{d-1}{d}\, \hat{h}(\Box-m^2)\hat{h}\]
\, .\label{SFPhath2}
\ee
We see that  $h^{\rm TT}_{\mu\nu}$ has a healthy kinetic term. In contrast, for $d>1$, in a theory governed by $S_{\rm non-loc}$
the scalar $\hat{h}$ is a ghost, since its kinetic term has the wrong sign. 
\Eq{SFPhath2} confirms the analysis made in Sect.~\ref{sect:dof}: the action
$S_{\rm non-loc}$ describes six degrees of freedom, out of which five correspond to the helicities $0,\pm 1$ and $\pm 2$ of a massive spin-2 particle, while the sixth degree of freedom is a Lorentz scalar, and we further see that it is a ghost.
Taking the $m=0$ limit and comparing with \eq{SEH2nl} we see that in this limit $\hat{h}$ reduces to the non-radiative field $s$.
However, from the discussion in sect.~\ref{sect:dof}  it follows that
in the massive case $\hat{h}$ is truly dynamical, while we saw that in the massless case $s$ is a non-radiative degree of freedom, despite its KG action. Similarly, in the massive case
$h^{\rm TT}_{\mu\nu}$ describes five dynamical degrees of freedom.

Consider now FP massive gravity. According to \eq{SmassiveGNonLoc}, we must then add
to $S_{\rm non-loc}$ the term
\be
-2m^2\int d^4x\, N\frac{1}{\Box}\pam\pan(\hMN-\eMN h)\, .
\ee
Observe, from \eq{defhath}, that 
\be
\hat{h}=-\frac{1}{\Box}\pam\pan (\hMN-\eMN h)\, .
\ee
Therefore the FP action (\ref{SmassiveGNonLoc}) can be rewritten as
\bees
S_{\rm FP}+S_{\rm int}&=&\int d^{d+1}x\, \[  
\frac{1}{2}h^{\rm TT}_{\mu\nu}(\Box-m^2)h^{{\rm TT}\mu\nu}
- \frac{d-1}{2d}\hat{h}(\Box-m^2)\hat{h}
+2m^2{N}\hat{h}\]\nn\\
&& + \frac{\kappa}{2}\int d^{d+1}x\,  \[ h^{\rm TT}_{\mu\nu}T^{{\rm TT}\mu\nu} +\frac{1}{d}\hat{h}T\]
\, ,\label{SFPhath}
\ees 
We also wrote \eq{SmassiveGNonLoc} in a generic space-time dimension and we used the fact that, because of $\pam\TMN=0$, upon integration by parts $\hmn\TMN=\hat{h}_{\mu\nu}\TMN = \hmn^{\rm TT}T^{{\rm TT}\mu\nu}
 +(1/d)\hat{h}T$.\footnote{Assuming that $\Tmn$ has compact support, the integrations by parts involving non-local terms are well defined.}
We see that
the Lagrange multiplier $N$ imposes the constraint $\hat{h}=0$ and kills the ghost.
Thus, the action 
(\ref {SFPhath}) is equivalent to
\be\label{S2intH}
S_{\rm FP}+S_{\rm int} = \int d^{d+1}x\, \[  \frac{1}{2}\hmn^{\rm TT}(\Box-m^2)h^{{\rm TT}\mu\nu}
+\frac{\kappa}{2}\hmn^{\rm TT}T^{{\rm TT}\mu\nu}\]\, .
\ee
\Eq {S2intH} provides a gauge-invariant description of the five physical degrees of freedom of the massive graviton of FP theory. The price to be paid for explicit gauge invariance is of course non-locality, which is now hidden in the relation between the gauge-invariant field $\hmn^{\rm TT}$ and the metric perturbation $\hmn$, given by \eqs {defhath} {defHmunu}. It is also interesting to observe that, under an infinitesimal conformal transformation of the metric $\gmn\ra e^{-2\theta}\gmn$, we have
$\hmn\ra \hmn-2\theta\emn$ and
\be
\hat{h}_{\mu\nu}\ra \hat{h}_{\mu\nu}-2 \(\emn-\iBox \pam\pan\)\theta\, .
\ee
Plugging this into \eq{defHmunu} we find that $\hmn^{\rm TT}$ is invariant. Thus, the reduction from the 10 degrees of freedom of $\hmn$ to the five of $\hmn^{\rm TT}$ can be understood as a consequence of diff invariance (which eliminates four degrees of freedom) plus an ``accidental" conformal invariance of the linearized theory.

\bibliographystyle{utphys}
\bibliography{myrefs_massive}

\end{document}

%% file: mydefs.tex
\newcommand{\iBox}{\Box^{-1}}
\newcommand{\Stu}{St\"uckelberg }

\newcommand{\Fmn}{F_{\mu\nu}}
\newcommand{\FMN}{F^{\mu\nu}}
\newcommand{\Am}{A_{\mu}}
\newcommand{\An}{A_{\nu}}
\newcommand{\AMU}{A^{\mu}}
\newcommand{\AN}{A^{\nu}}

\renewcommand\({\left(}
\renewcommand\){\right)}
\renewcommand\[{\left[}
\renewcommand\]{\right]}

\newcommand\n{{\mbox {\boldmath $\nabla$}}}
\newcommand{\ra}{\rightarrow}

\def\lsim{\raise 0.4ex\hbox{$<$}\kern -0.8em\lower 0.62
ex\hbox{$\sim$}}

\def\gsim{\raise 0.4ex\hbox{$>$}\kern -0.7em\lower 0.62
ex\hbox{$\sim$}}

\def\lbar{{\hbox{$\lambda$}\kern -0.7em\raise 0.6ex
\hbox{$-$}}}

\newcommand\eq[1]{eq.~(\ref{#1})}
\newcommand\eqs[2]{eqs.~(\ref{#1}) and (\ref{#2})}
\newcommand\Eq[1]{Equation~(\ref{#1})}
\newcommand\Eqs[2]{Equations~(\ref{#1}) and (\ref{#2})}

\newcommand\eqst[2]{eqs.~(\ref{#1})--(\ref{#2})}

\newcommand\pa{\partial}
\newcommand\p{\partial}

\newcommand\ee{\end{equation}}
\newcommand\be{\begin{equation}}
\def\bea{\begin{array}}
\def\eea{\end{array}}\def\ea{\end{array}}
\newcommand\ees{\end{eqnarray}}
\newcommand\bees{\begin{eqnarray}}
\def\nn{\nonumber}

\def\a{\alpha}

\def\s{\sigma}
\def\g{\gamma}

\def\d{\delta}

\def\eps{\epsilon}

\def\dslash{\hspace{-1mm}\not{\hbox{\kern-2pt $\partial$}}}
\def\Dslash{\not{\hbox{\kern-4pt $D$}}}
\def\pslash{\not{\hbox{\kern-2.1pt $p$}}}
\def\kslash{\not{\hbox{\kern-2.3pt $k$}}}
\def\qslash{\not{\hbox{\kern-2.3pt $q$}}}

\newcommand{\vx}{{\bf x}}

\def\p1{{\bf p}_1}
\def\p2{{\bf p}_2}
\def\k1{{\bf k}_1}
\def\k2{{\bf k}_2}

\newcommand{\emn}{\eta_{\mu\nu}}
\newcommand{\ers}{\eta_{\rho\sigma}}
\newcommand{\emr}{\eta_{\mu\rho}}
\newcommand{\ens}{\eta_{\nu\sigma}}
\newcommand{\ems}{\eta_{\mu\sigma}}
\newcommand{\enr}{\eta_{\nu\rho}}
\newcommand{\eMN}{\eta^{\mu\nu}}
\newcommand{\eRS}{\eta^{\rho\sigma}}
\newcommand{\eMR}{\eta^{\mu\rho}}
\newcommand{\eNS}{\eta^{\nu\sigma}}
\newcommand{\eMS}{\eta^{\mu\sigma}}
\newcommand{\eNR}{\eta^{\nu\rho}}

\newcommand{\gmn}{g_{\mu\nu}}

\newcommand{\gMN}{g^{\mu\nu}}

\newcommand{\gbmn}{\bar{g}_{\mu\nu}}

\newcommand{\hmn}{h_{\mu\nu}}
\newcommand{\hrs}{h_{\rho\sigma}}
\newcommand{\hmr}{h_{\mu\rho}}

\newcommand{\hnr}{h_{\nu\rho}}

\newcommand{\hMN}{h^{\mu\nu}}
\newcommand{\hRS}{h^{\rho\sigma}}
\newcommand{\hMR}{h^{\mu\rho}}

\newcommand{\hhmn}{\hat{h}_{\mu\nu}}

\newcommand{\bhmn}{\bar{h}_{\mu\nu}}

\newcommand{\bhMN}{\bar{h}^{\mu\nu}}

\newcommand{\xm}{x^{\mu}}

\newcommand{\xim}{\xi_{\mu}}
\newcommand{\xin}{\xi_{\nu}}

\newcommand{\xiM}{\xi^{\mu}}

\newcommand{\pam}{\pa_{\mu}}

\newcommand{\pan}{\pa_{\nu}}
\newcommand{\parho}{\pa_{\rho}}
\newcommand{\pas}{\pa_{\sigma}}
\newcommand{\paM}{\pa^{\mu}}
\newcommand{\paN}{\pa^{\nu}}
\newcommand{\paR}{\pa^{\rho}}
\newcommand{\paS}{\pa^{\sigma}}

\newcommand{\Gmn}{G_{\mu\nu}}

\newcommand{\GMN}{G^{\mu\nu}}

\newcommand{\Tmn}{T_{\mu\nu}}

\newcommand{\TMN}{T^{\mu\nu}}

\newcommand{\dddM}{\kern 0.2em \raise 1.9ex\hbox{$...$}\kern -1.0em \hbox{$M$}}
\newcommand{\dddQ}{\kern 0.2em \raise 1.9ex\hbox{$...$}\kern -1.0em \hbox{$Q$}}
\newcommand{\dddI}{\kern 0.2em \raise 1.9ex\hbox{$...$}\kern -1.0em\hbox{$I$}}
\newcommand{\dddJ}{\kern 0.2em \raise 1.9ex\hbox{$...$}\kern-1.0em
\hbox{$J$}}
\newcommand{\dddcalJ}{\kern 0.2em \raise 1.9ex\hbox{$...$}\kern-1.0em
\hbox{${\cal J}$}}

\newcommand{\dddO}{\kern 0.2em \raise 1.9ex\hbox{$...$}\kern -1.0em
\hbox{${\cal O}$}}
\def\dddz{\raise 1.5ex\hbox{$...$}\kern -0.8em \hbox{$z$}}
\def\dddd{\raise 1.8ex\hbox{$...$}\kern -0.8em \hbox{$d$}}
\def\dddbd{\raise 1.8ex\hbox{$...$}\kern -0.8em \hbox{${\bf d}$}}
\def\ddbd{\raise 1.8ex\hbox{$..$}\kern -0.8em \hbox{${\bf d}$}}
\def\dddx{\raise 1.6ex\hbox{$...$}\kern -0.8em \hbox{$x$}}

\newcommand{\mpl}{M_{\rm Pl}}






%% file: paper160713.bbl
\providecommand{\href}[2]{#2}\begingroup\raggedright\begin{thebibliography}{100}

\bibitem{Fierz:1939ix}
M.~Fierz and W.~Pauli, ``{On relativistic wave equations for particles of
  arbitrary spin in an electromagnetic field},'' {\em Proc.Roy.Soc.Lond.} {\bf
  A173} (1939) 211--232.

\bibitem{Boulware:1973my}
D.~Boulware and S.~Deser, ``{Can gravitation have a finite range?},'' {\em
  Phys.Rev.} {\bf D6} (1972) 3368--3382.

\bibitem{Dvali:2000hr}
G.~Dvali, G.~Gabadadze, and M.~Porrati, ``{4-D gravity on a brane in 5-D
  Minkowski space},'' {\em Phys.Lett.} {\bf B485} (2000) 208--214,
  \href{http://xxx.lanl.gov/abs/hep-th/0005016}{{\tt hep-th/0005016}}.

\bibitem{Dvali:2000xg}
G.~Dvali and G.~Gabadadze, ``{Gravity on a brane in infinite volume extra
  space},'' {\em Phys.Rev.} {\bf D63} (2001) 065007,
  \href{http://xxx.lanl.gov/abs/hep-th/0008054}{{\tt hep-th/0008054}}.

\bibitem{Dvali:2002pe}
G.~Dvali, G.~Gabadadze, and M.~Shifman, ``{Diluting cosmological constant in
  infinite volume extra dimensions},'' {\em Phys.Rev.} {\bf D67} (2003) 044020,
  \href{http://xxx.lanl.gov/abs/hep-th/0202174}{{\tt hep-th/0202174}}.

\bibitem{ArkaniHamed:2002fu}
N.~Arkani-Hamed, S.~Dimopoulos, G.~Dvali, and G.~Gabadadze, ``{Nonlocal
  modification of gravity and the cosmological constant problem},''
  \href{http://xxx.lanl.gov/abs/hep-th/0209227}{{\tt hep-th/0209227}}.

\bibitem{Dvali:2006su}
G.~Dvali, ``{Predictive Power of Strong Coupling in Theories with Large
  Distance Modified Gravity},'' {\em New J.Phys.} {\bf 8} (2006) 326,
  \href{http://xxx.lanl.gov/abs/hep-th/0610013}{{\tt hep-th/0610013}}.

\bibitem{Dvali:2007kt}
G.~Dvali, S.~Hofmann, and J.~Khoury, ``{Degravitation of the cosmological
  constant and graviton width},'' {\em Phys.Rev.} {\bf D76} (2007) 084006,
  \href{http://xxx.lanl.gov/abs/hep-th/0703027}{{\tt hep-th/0703027}}.

\bibitem{ArkaniHamed:2002sp}
N.~Arkani-Hamed, H.~Georgi, and M.~D. Schwartz, ``{Effective field theory for
  massive gravitons and gravity in theory space},'' {\em Annals Phys.} {\bf
  305} (2003) 96--118, \href{http://xxx.lanl.gov/abs/hep-th/0210184}{{\tt
  hep-th/0210184}}.

\bibitem{Nicolis:2008in}
A.~Nicolis, R.~Rattazzi, and E.~Trincherini, ``{The Galileon as a local
  modification of gravity},'' {\em Phys.Rev.} {\bf D79} (2009) 064036,
  \href{http://xxx.lanl.gov/abs/0811.2197}{{\tt 0811.2197}}.

\bibitem{deRham:2010ik}
C.~de~Rham and G.~Gabadadze, ``{Generalization of the Fierz-Pauli Action},''
  {\em Phys.Rev.} {\bf D82} (2010) 044020,
  \href{http://xxx.lanl.gov/abs/1007.0443}{{\tt 1007.0443}}.

\bibitem{deRham:2010kj}
C.~de~Rham, G.~Gabadadze, and A.~J. Tolley, ``{Resummation of Massive
  Gravity},'' {\em Phys.Rev.Lett.} {\bf 106} (2011) 231101,
  \href{http://xxx.lanl.gov/abs/1011.1232}{{\tt 1011.1232}}.

\bibitem{deRham:2010gu}
C.~de~Rham and G.~Gabadadze, ``{Selftuned Massive Spin-2},'' {\em Phys.Lett.}
  {\bf B693} (2010) 334--338, \href{http://xxx.lanl.gov/abs/1006.4367}{{\tt
  1006.4367}}.

\bibitem{deRham:2011rn}
C.~de~Rham, G.~Gabadadze, and A.~J. Tolley, ``{Ghost free Massive Gravity in
  the St\'uckelberg language},'' {\em Phys.Lett.} {\bf B711} (2012) 190--195,
  \href{http://xxx.lanl.gov/abs/1107.3820}{{\tt 1107.3820}}.

\bibitem{deRham:2011qq}
C.~de~Rham, G.~Gabadadze, and A.~J. Tolley, ``{Helicity Decomposition of
  Ghost-free Massive Gravity},'' {\em JHEP} {\bf 1111} (2011) 093,
  \href{http://xxx.lanl.gov/abs/1108.4521}{{\tt 1108.4521}}.

\bibitem{Hassan:2011hr}
S.~Hassan and R.~A. Rosen, ``{Resolving the Ghost Problem in non-Linear Massive
  Gravity},'' {\em Phys.Rev.Lett.} {\bf 108} (2012) 041101,
  \href{http://xxx.lanl.gov/abs/1106.3344}{{\tt 1106.3344}}.

\bibitem{Hassan:2011vm}
S.~Hassan and R.~A. Rosen, ``{On Non-Linear Actions for Massive Gravity},''
  {\em JHEP} {\bf 1107} (2011) 009,
  \href{http://xxx.lanl.gov/abs/1103.6055}{{\tt 1103.6055}}.

\bibitem{Hassan:2011tf}
S.~Hassan, R.~A. Rosen, and A.~Schmidt-May, ``{Ghost-free Massive Gravity with
  a General Reference Metric},'' {\em JHEP} {\bf 1202} (2012) 026,
  \href{http://xxx.lanl.gov/abs/1109.3230}{{\tt 1109.3230}}.

\bibitem{Hassan:2011ea}
S.~Hassan and R.~A. Rosen, ``{Confirmation of the Secondary Constraint and
  Absence of Ghost in Massive Gravity and Bimetric Gravity},'' {\em JHEP} {\bf
  1204} (2012) 123, \href{http://xxx.lanl.gov/abs/1111.2070}{{\tt 1111.2070}}.

\bibitem{Hassan:2012qv}
S.~Hassan, A.~Schmidt-May, and M.~von Strauss, ``{Proof of Consistency of
  Nonlinear Massive Gravity in the St\"uckelberg Formulation},'' {\em
  Phys.Lett.} {\bf B715} (2012) 335--339,
  \href{http://xxx.lanl.gov/abs/1203.5283}{{\tt 1203.5283}}.

\bibitem{Comelli:2012vz}
D.~Comelli, M.~Crisostomi, F.~Nesti, and L.~Pilo, ``{Degrees of Freedom in
  Massive Gravity},'' {\em Phys.Rev.} {\bf D86} (2012) 101502,
  \href{http://xxx.lanl.gov/abs/1204.1027}{{\tt 1204.1027}}.

\bibitem{Comelli:2013txa}
D.~Comelli, F.~Nesti, and L.~Pilo, ``{Massive gravity: a General Analysis},''
  \href{http://xxx.lanl.gov/abs/1305.0236}{{\tt 1305.0236}}.

\bibitem{Hinterbichler:2011tt}
K.~Hinterbichler, ``{Theoretical Aspects of Massive Gravity},'' {\em
  Rev.Mod.Phys.} {\bf 84} (2012) 671--710,
  \href{http://xxx.lanl.gov/abs/1105.3735}{{\tt 1105.3735}}.

\bibitem{Deffayet:2000uy}
C.~Deffayet, ``{Cosmology on a brane in Minkowski bulk},'' {\em Phys.Lett.}
  {\bf B502} (2001) 199--208,
  \href{http://xxx.lanl.gov/abs/hep-th/0010186}{{\tt hep-th/0010186}}.

\bibitem{Deffayet:2001pu}
C.~Deffayet, G.~Dvali, and G.~Gabadadze, ``{Accelerated universe from gravity
  leaking to extra dimensions},'' {\em Phys.Rev.} {\bf D65} (2002) 044023,
  \href{http://xxx.lanl.gov/abs/astro-ph/0105068}{{\tt astro-ph/0105068}}.

\bibitem{Luty:2003vm}
M.~A. Luty, M.~Porrati, and R.~Rattazzi, ``{Strong interactions and stability
  in the DGP model},'' {\em JHEP} {\bf 0309} (2003) 029,
  \href{http://xxx.lanl.gov/abs/hep-th/0303116}{{\tt hep-th/0303116}}.

\bibitem{Nicolis:2004qq}
A.~Nicolis and R.~Rattazzi, ``{Classical and quantum consistency of the DGP
  model},'' {\em JHEP} {\bf 0406} (2004) 059,
  \href{http://xxx.lanl.gov/abs/hep-th/0404159}{{\tt hep-th/0404159}}.

\bibitem{Gorbunov:2005zk}
D.~Gorbunov, K.~Koyama, and S.~Sibiryakov, ``{More on ghosts in DGP model},''
  {\em Phys.Rev.} {\bf D73} (2006) 044016,
  \href{http://xxx.lanl.gov/abs/hep-th/0512097}{{\tt hep-th/0512097}}.

\bibitem{Charmousis:2006pn}
C.~Charmousis, R.~Gregory, N.~Kaloper, and A.~Padilla, ``{DGP Specteroscopy},''
  {\em JHEP} {\bf 0610} (2006) 066,
  \href{http://xxx.lanl.gov/abs/hep-th/0604086}{{\tt hep-th/0604086}}.

\bibitem{Izumi:2006ca}
K.~Izumi, K.~Koyama, and T.~Tanaka, ``{Unexorcized ghost in DGP brane world},''
  {\em JHEP} {\bf 0704} (2007) 053,
  \href{http://xxx.lanl.gov/abs/hep-th/0610282}{{\tt hep-th/0610282}}.

\bibitem{deRham:2010tw}
C.~de~Rham, G.~Gabadadze, L.~Heisenberg, and D.~Pirtskhalava, ``{Cosmic
  Acceleration and the Helicity-0 Graviton},'' {\em Phys.Rev.} {\bf D83} (2011)
  103516, \href{http://xxx.lanl.gov/abs/1010.1780}{{\tt 1010.1780}}.

\bibitem{Koyama:2011xz}
K.~Koyama, G.~Niz, and G.~Tasinato, ``{Analytic solutions in non-linear massive
  gravity},'' {\em Phys.Rev.Lett.} {\bf 107} (2011) 131101,
  \href{http://xxx.lanl.gov/abs/1103.4708}{{\tt 1103.4708}}.

\bibitem{Koyama:2011yg}
K.~Koyama, G.~Niz, and G.~Tasinato, ``{Strong interactions and exact solutions
  in non-linear massive gravity},'' {\em Phys.Rev.} {\bf D84} (2011) 064033,
  \href{http://xxx.lanl.gov/abs/1104.2143}{{\tt 1104.2143}}.

\bibitem{Nieuwenhuizen:2011sq}
T.~Nieuwenhuizen, ``{Exact Schwarzschild-de Sitter black holes in a family of
  massive gravity models},'' {\em Phys.Rev.} {\bf D84} (2011) 024038,
  \href{http://xxx.lanl.gov/abs/1103.5912}{{\tt 1103.5912}}.

\bibitem{Chamseddine:2011bu}
A.~H. Chamseddine and M.~S. Volkov, ``{Cosmological solutions with massive
  gravitons},'' {\em Phys.Lett.} {\bf B704} (2011) 652--654,
  \href{http://xxx.lanl.gov/abs/1107.5504}{{\tt 1107.5504}}.

\bibitem{D'Amico:2011jj}
G.~D'Amico, C.~de~Rham, S.~Dubovsky, G.~Gabadadze, D.~Pirtskhalava, and A.~J.
  Tolley, ``{Massive Cosmologies},'' {\em Phys.Rev.} {\bf D84} (2011) 124046,
  \href{http://xxx.lanl.gov/abs/1108.5231}{{\tt 1108.5231}}.

\bibitem{DeFelice:2013bxa}
A.~De~Felice, A.~E. Gumrukcuoglu, C.~Lin, and S.~Mukohyama, ``{On the cosmology
  of massive gravity},'' \href{http://xxx.lanl.gov/abs/1304.0484}{{\tt
  1304.0484}}.

\bibitem{Tasinato:2013rza}
G.~Tasinato, K.~Koyama, and G.~Niz, ``{Exact Solutions in Massive Gravity},''
  \href{http://xxx.lanl.gov/abs/1304.0601}{{\tt 1304.0601}}.

\bibitem{Gumrukcuoglu:2011ew}
A.~E. Gumrukcuoglu, C.~Lin, and S.~Mukohyama, ``{Open FRW universes and
  self-acceleration from nonlinear massive gravity},'' {\em JCAP} {\bf 1111}
  (2011) 030, \href{http://xxx.lanl.gov/abs/1109.3845}{{\tt 1109.3845}}.

\bibitem{Gumrukcuoglu:2011zh}
A.~E. Gumrukcuoglu, C.~Lin, and S.~Mukohyama, ``{Cosmological perturbations of
  self-accelerating universe in nonlinear massive gravity},'' {\em JCAP} {\bf
  1203} (2012) 006, \href{http://xxx.lanl.gov/abs/1111.4107}{{\tt 1111.4107}}.

\bibitem{Koyama:2011wx}
K.~Koyama, G.~Niz, and G.~Tasinato, ``{The Self-Accelerating Universe with
  Vectors in Massive Gravity},'' {\em JHEP} {\bf 1112} (2011) 065,
  \href{http://xxx.lanl.gov/abs/1110.2618}{{\tt 1110.2618}}.

\bibitem{Dubovsky:2005xd}
S.~Dubovsky, T.~Gregoire, A.~Nicolis, and R.~Rattazzi, ``{Null energy condition
  and superluminal propagation},'' {\em JHEP} {\bf 0603} (2006) 025,
  \href{http://xxx.lanl.gov/abs/hep-th/0512260}{{\tt hep-th/0512260}}.

\bibitem{Gruzinov:2011sq}
A.~Gruzinov, ``{All Fierz-Paulian massive gravity theories have ghosts or
  superluminal modes},'' \href{http://xxx.lanl.gov/abs/1106.3972}{{\tt
  1106.3972}}.

\bibitem{deRham:2011pt}
C.~de~Rham, G.~Gabadadze, and A.~J. Tolley, ``{Comments on
  (super)luminality},'' \href{http://xxx.lanl.gov/abs/1107.0710}{{\tt
  1107.0710}}.

\bibitem{Deser:2012qx}
S.~Deser and A.~Waldron, ``{Acausality of Massive Gravity},'' {\em Phys. Rev.
  Lett. 110,} {\bf 111101} (2013) \href{http://xxx.lanl.gov/abs/1212.5835}{{\tt
  1212.5835}}.

\bibitem{Berezhiani:2013dw}
L.~Berezhiani, G.~Chkareuli, and G.~Gabadadze, ``{Restricted Galileons},''
  \href{http://xxx.lanl.gov/abs/1302.0549}{{\tt 1302.0549}}.

\bibitem{Izumi:2013poa}
K.~Izumi and Y.~C. Ong, ``{An Analysis of Characteristics in Non-Linear Massive
  Gravity},'' \href{http://xxx.lanl.gov/abs/1304.0211}{{\tt 1304.0211}}.

\bibitem{Berezhiani:2013dca}
L.~Berezhiani, G.~Chkareuli, C.~de~Rham, G.~Gabadadze, and A.~Tolley, ``{Mixed
  Galileons and Spherically Symmetric Solutions},''
  \href{http://xxx.lanl.gov/abs/1305.0271}{{\tt 1305.0271}}.

\bibitem{Koyama:2013paa}
K.~Koyama, G.~Niz, and G.~Tasinato, ``{Effective theory for the Vainshtein
  mechanism from the Horndeski action},''
  \href{http://xxx.lanl.gov/abs/1305.0279}{{\tt 1305.0279}}.

\bibitem{Nicolis:2009qm}
A.~Nicolis, R.~Rattazzi, and E.~Trincherini, ``{Energy's and amplitudes'
  positivity},'' {\em JHEP} {\bf 1005} (2010) 095,
  \href{http://xxx.lanl.gov/abs/0912.4258}{{\tt 0912.4258}}.

\bibitem{Burrage:2011cr}
C.~Burrage, C.~de~Rham, L.~Heisenberg, and A.~J. Tolley, ``{Chronology
  Protection in Galileon Models and Massive Gravity},'' {\em JCAP} {\bf 1207}
  (2012) 004, \href{http://xxx.lanl.gov/abs/1111.5549}{{\tt 1111.5549}}.

\bibitem{Hassan:2011zd}
S.~Hassan and R.~A. Rosen, ``{Bimetric Gravity from Ghost-free Massive
  Gravity},'' {\em JHEP} {\bf 1202} (2012) 126,
  \href{http://xxx.lanl.gov/abs/1109.3515}{{\tt 1109.3515}}.

\bibitem{Hassan:2012wr}
S.~Hassan, A.~Schmidt-May, and M.~von Strauss, ``{On Consistent Theories of
  Massive Spin-2 Fields Coupled to Gravity},''
  \href{http://xxx.lanl.gov/abs/1208.1515}{{\tt 1208.1515}}.

\bibitem{Hassan:2012gz}
S.~Hassan, A.~Schmidt-May, and M.~von Strauss, ``{On Partially Massless
  Bimetric Gravity},'' \href{http://xxx.lanl.gov/abs/1208.1797}{{\tt
  1208.1797}}.

\bibitem{Comelli:2012db}
D.~Comelli, M.~Crisostomi, and L.~Pilo, ``{Perturbations in Massive Gravity
  Cosmology},'' {\em JHEP} {\bf 1206} (2012) 085,
  \href{http://xxx.lanl.gov/abs/1202.1986}{{\tt 1202.1986}}.

\bibitem{Berg:2012kn}
M.~Berg, I.~Buchberger, J.~Enander, E.~Mortsell, and S.~Sjors, ``{Growth
  Histories in Bimetric Massive Gravity},''
  \href{http://xxx.lanl.gov/abs/1206.3496}{{\tt 1206.3496}}.

\bibitem{Eliezer:1989cr}
D.~Eliezer and R.~Woodard, ``{The Problem of Nonlocality in String Theory},''
  {\em Nucl.Phys.} {\bf B325} (1989) 389.

\bibitem{Simon:1990ic}
J.~Z. Simon, ``{Higher derivative lagrangians, nonlocality, problems and
  solutions},'' {\em Phys.Rev.} {\bf D41} (1990) 3720.

\bibitem{Woodard:2006nt}
R.~P. Woodard, ``{Avoiding dark energy with 1/R modifications of gravity},''
  {\em Lect.Notes Phys.} {\bf 720} (2007) 403--433,
  \href{http://xxx.lanl.gov/abs/astro-ph/0601672}{{\tt astro-ph/0601672}}.

\bibitem{Barnaby:2007ve}
N.~Barnaby and N.~Kamran, ``{Dynamics with infinitely many derivatives: The
  Initial value problem},'' {\em JHEP} {\bf 0802} (2008) 008,
  \href{http://xxx.lanl.gov/abs/0709.3968}{{\tt 0709.3968}}.

\bibitem{Barvinsky:1985an}
A.~Barvinsky and G.~Vilkovisky, ``{The Generalized Schwinger-Dewitt Technique
  in Gauge Theories and Quantum Gravity},'' {\em Phys.Rept.} {\bf 119} (1985)
  1--74.

\bibitem{Barvinsky:1987uw}
A.~Barvinsky and G.~Vilkovisky, ``{Beyond the Schwinger-Dewitt Technique:
  Converting Loops Into Trees and In-In Currents},'' {\em Nucl.Phys.} {\bf
  B282} (1987) 163--188.

\bibitem{Ostrovsky:1988jn}
A.~Ostrovsky and G.~Vilkovisky, ``{The covariant effective action in QED. One
  loop magnetic moment},'' {\em J.Math.Phys.} {\bf 29} (1988) 702--708.

\bibitem{Buchbinder:1992rb}
I.~L. Buchbinder, S.~D. Odintsov, and I.~L. Shapiro, {\em {Effective action in
  quantum gravity}}.
\newblock Institute of Physics, Bristol, UK, 1992.

\bibitem{Dobado:1998mr}
A.~Dobado and A.~L. Maroto, ``{Particle production from nonlocal gravitational
  effective action},'' {\em Phys.Rev.} {\bf D60} (1999) 104045,
  \href{http://xxx.lanl.gov/abs/gr-qc/9803076}{{\tt gr-qc/9803076}}.

\bibitem{Barvinsky:2003kg}
A.~Barvinsky, ``{Nonlocal action for long distance modifications of gravity
  theory},'' {\em Phys.Lett.} {\bf B572} (2003) 109--116,
  \href{http://xxx.lanl.gov/abs/hep-th/0304229}{{\tt hep-th/0304229}}.

\bibitem{Shapiro:2008sf}
I.~L. Shapiro, ``{Effective Action of Vacuum: Semiclassical Approach},'' {\em
  Class. Quant. Grav.} {\bf 25} (2008) 103001,
  \href{http://xxx.lanl.gov/abs/0801.0216}{{\tt 0801.0216}}.

\bibitem{Shapiro:2009dh}
I.~L. Shapiro and J.~Sola, ``{On the possible running of the cosmological
  'constant'},'' {\em Phys.Lett.} {\bf B682} (2009) 105--113,
  \href{http://xxx.lanl.gov/abs/0910.4925}{{\tt 0910.4925}}.

\bibitem{Hamber:2005dw}
H.~Hamber and R.~M. Williams, ``{Nonlocal effective gravitational field
  equations and the running of Newton's G},'' {\em Phys.Rev.} {\bf D72} (2005)
  044026, \href{http://xxx.lanl.gov/abs/hep-th/0507017}{{\tt hep-th/0507017}}.

\bibitem{Khoury:2006fg}
J.~Khoury, ``{Fading gravity and self-inflation},'' {\em Phys.Rev.} {\bf D76}
  (2007) 123513, \href{http://xxx.lanl.gov/abs/hep-th/0612052}{{\tt
  hep-th/0612052}}.

\bibitem{Modesto:2011kw}
L.~Modesto, ``{Super-renormalizable Quantum Gravity},'' {\em Phys.Rev.} {\bf
  D86} (2012) 044005, \href{http://xxx.lanl.gov/abs/1107.2403}{{\tt
  1107.2403}}.

\bibitem{Briscese:2012ys}
F.~Briscese, A.~Marciano, L.~Modesto, and E.~N. Saridakis, ``{Inflation in
  (Super-)renormalizable Gravity},'' {\em Phys.Rev.} {\bf D87} (2013) 083507,
  \href{http://xxx.lanl.gov/abs/1212.3611}{{\tt 1212.3611}}.

\bibitem{Francia:2002aa}
D.~Francia and A.~Sagnotti, ``{Free geometric equations for higher spins},''
  {\em Phys.Lett.} {\bf B543} (2002) 303--310,
  \href{http://xxx.lanl.gov/abs/hep-th/0207002}{{\tt hep-th/0207002}}.

\bibitem{Porrati:2002cp}
M.~Porrati, ``{Fully covariant van Dam-Veltman-Zakharov discontinuity, and
  absence thereof},'' {\em Phys.Lett.} {\bf B534} (2002) 209--215,
  \href{http://xxx.lanl.gov/abs/hep-th/0203014}{{\tt hep-th/0203014}}.

\bibitem{Vainshtein:1972sx}
A.~Vainshtein, ``{To the problem of nonvanishing gravitation mass},'' {\em
  Phys.Lett.} {\bf B39} (1972) 393--394.

\bibitem{Deser:2007jk}
S.~Deser and R.~Woodard, ``{Nonlocal Cosmology},'' {\em Phys.Rev.Lett.} {\bf
  99} (2007) 111301, \href{http://xxx.lanl.gov/abs/0706.2151}{{\tt 0706.2151}}.

\bibitem{Jhingan:2008ym}
S.~Jhingan, S.~Nojiri, S.~Odintsov, M.~Sami, I.~Thongkool, {\em et.~al.},
  ``{Phantom and non-phantom dark energy: The Cosmological relevance of
  non-locally corrected gravity},'' {\em Phys.Lett.} {\bf B663} (2008)
  424--428, \href{http://xxx.lanl.gov/abs/0803.2613}{{\tt 0803.2613}}.

\bibitem{Koivisto:2008xfa}
T.~Koivisto, ``{Dynamics of Nonlocal Cosmology},'' {\em Phys.Rev.} {\bf D77}
  (2008) 123513, \href{http://xxx.lanl.gov/abs/0803.3399}{{\tt 0803.3399}}.

\bibitem{Koivisto:2008dh}
T.~Koivisto, ``{Newtonian limit of nonlocal cosmology},'' {\em Phys.Rev.} {\bf
  D78} (2008) 123505, \href{http://xxx.lanl.gov/abs/0807.3778}{{\tt
  0807.3778}}.

\bibitem{Capozziello:2008gu}
S.~Capozziello, E.~Elizalde, S.~Nojiri, and S.~D. Odintsov, ``{Accelerating
  cosmologies from non-local higher-derivative gravity},'' {\em Phys.Lett.}
  {\bf B671} (2009) 193--198, \href{http://xxx.lanl.gov/abs/0809.1535}{{\tt
  0809.1535}}.

\bibitem{Nojiri:2010pw}
S.~Nojiri, S.~D. Odintsov, M.~Sasaki, and Y.~Zhang, ``{Screening of
  cosmological constant in non-local gravity},'' {\em Phys.Lett.} {\bf B696}
  (2011) 278--282, \href{http://xxx.lanl.gov/abs/1010.5375}{{\tt 1010.5375}}.

\bibitem{Bamba:2012ky}
K.~Bamba, S.~Nojiri, S.~D. Odintsov, and M.~Sasaki, ``{Screening of
  cosmological constant for De Sitter Universe in non-local gravity,
  phantom-divide crossing and finite-time future singularities},'' {\em
  Gen.Rel.Grav.} {\bf 44} (2012) 1321--1356,
  \href{http://xxx.lanl.gov/abs/1104.2692}{{\tt 1104.2692}}.

\bibitem{Kluson:2011tb}
J.~Kluson, ``{Non-Local Gravity from Hamiltonian Point of View},'' {\em JHEP}
  {\bf 1109} (2011) 001, \href{http://xxx.lanl.gov/abs/1105.6056}{{\tt
  1105.6056}}.

\bibitem{Barvinsky:2011hd}
A.~Barvinsky, ``{Dark energy and dark matter from nonlocal ghost-free gravity
  theory},'' {\em Phys.Lett.} {\bf B710} (2012) 12--16,
  \href{http://xxx.lanl.gov/abs/1107.1463}{{\tt 1107.1463}}.

\bibitem{Zhang:2011uv}
Y.~Zhang and M.~Sasaki, ``{Screening of cosmological constant in non-local
  cosmology},'' {\em Int.J.Mod.Phys.} {\bf D21} (2012) 1250006,
  \href{http://xxx.lanl.gov/abs/1108.2112}{{\tt 1108.2112}}.

\bibitem{Elizalde:2011su}
E.~Elizalde, E.~Pozdeeva, and S.~Y. Vernov, ``{De Sitter Universe in Non-local
  Gravity},'' {\em Phys.Rev.} {\bf D85} (2012) 044002,
  \href{http://xxx.lanl.gov/abs/1110.5806}{{\tt 1110.5806}}.

\bibitem{Park:2012cp}
S.~Park and S.~Dodelson, ``{Structure formation in a nonlocally modified
  gravity model},'' {\em Phys.Rev.} {\bf D87} (2013) 024003,
  \href{http://xxx.lanl.gov/abs/1209.0836}{{\tt 1209.0836}}.

\bibitem{Deffayet:2009ca}
C.~Deffayet and R.~Woodard, ``{Reconstructing the Distortion Function for
  Nonlocal Cosmology},'' {\em JCAP} {\bf 0908} (2009) 023,
  \href{http://xxx.lanl.gov/abs/0904.0961}{{\tt 0904.0961}}.

\bibitem{Elizalde:2012ja}
E.~Elizalde, E.~Pozdeeva, and S.~Y. Vernov, ``{Reconstruction Procedure in
  Nonlocal Models},'' {\em Class.Quant.Grav.} {\bf 30} (2013) 035002,
  \href{http://xxx.lanl.gov/abs/1209.5957}{{\tt 1209.5957}}.

\bibitem{Barvinsky:2011rk}
A.~O. Barvinsky, ``{Serendipitous discoveries in nonlocal gravity theory},''
  {\em Phys.Rev.} {\bf D85} (2012) 104018,
  \href{http://xxx.lanl.gov/abs/1112.4340}{{\tt 1112.4340}}.

\bibitem{Soussa:2003vv}
M.~Soussa and R.~P. Woodard, ``{A Nonlocal metric formulation of MOND},'' {\em
  Class.Quant.Grav.} {\bf 20} (2003) 2737--2752,
  \href{http://xxx.lanl.gov/abs/astro-ph/0302030}{{\tt astro-ph/0302030}}.

\bibitem{Nojiri:2007uq}
S.~Nojiri and S.~D. Odintsov, ``{Modified non-local-F(R) gravity as the key for
  the inflation and dark energy},'' {\em Phys.Lett.} {\bf B659} (2008)
  821--826, \href{http://xxx.lanl.gov/abs/0708.0924}{{\tt 0708.0924}}.

\bibitem{Babichev:2007dw}
E.~Babichev, V.~Mukhanov, and A.~Vikman, ``{k-Essence, superluminal
  propagation, causality and emergent geometry},'' {\em JHEP} {\bf 0802} (2008)
  101, \href{http://xxx.lanl.gov/abs/0708.0561}{{\tt 0708.0561}}.

\bibitem{Carroll:2003st}
S.~M. Carroll, M.~Hoffman, and M.~Trodden, ``{Can the dark energy
  equation-of-state parameter $w$ be less than -1?},'' {\em Phys.Rev.} {\bf
  D68} (2003) 023509, \href{http://xxx.lanl.gov/abs/astro-ph/0301273}{{\tt
  astro-ph/0301273}}.

\bibitem{Cline:2003gs}
J.~M. Cline, S.~Jeon, and G.~D. Moore, ``{The Phantom menaced: Constraints on
  low-energy effective ghosts},'' {\em Phys.Rev.} {\bf D70} (2004) 043543,
  \href{http://xxx.lanl.gov/abs/hep-ph/0311312}{{\tt hep-ph/0311312}}.

\bibitem{Kaplan:2005rr}
D.~E. Kaplan and R.~Sundrum, ``{A Symmetry for the cosmological constant},''
  {\em JHEP} {\bf 0607} (2006) 042,
  \href{http://xxx.lanl.gov/abs/hep-th/0505265}{{\tt hep-th/0505265}}.

\bibitem{Garriga:2012pk}
J.~Garriga and A.~Vilenkin, ``{Living with ghosts in Lorentz invariant
  theories},'' {\em JCAP} {\bf 1301} (2013) 036,
  \href{http://xxx.lanl.gov/abs/1202.1239}{{\tt 1202.1239}}.

\bibitem{Creminelli:2005qk}
P.~Creminelli, A.~Nicolis, M.~Papucci, and E.~Trincherini, ``{Ghosts in massive
  gravity},'' {\em JHEP} {\bf 0509} (2005) 003,
  \href{http://xxx.lanl.gov/abs/hep-th/0505147}{{\tt hep-th/0505147}}.

\bibitem{Gabadadze:2003jq}
G.~Gabadadze and A.~Gruzinov, ``{Graviton mass or cosmological constant?},''
  {\em Phys.Rev.} {\bf D72} (2005) 124007,
  \href{http://xxx.lanl.gov/abs/hep-th/0312074}{{\tt hep-th/0312074}}.

\bibitem{Myers:2003fd}
R.~C. Myers and M.~Pospelov, ``{Ultraviolet modifications of dispersion
  relations in effective field theory},'' {\em Phys.Rev.Lett.} {\bf 90} (2003)
  211601, \href{http://xxx.lanl.gov/abs/hep-ph/0301124}{{\tt hep-ph/0301124}}.

\bibitem{Flanagan:2005yc}
E.~E. Flanagan and S.~A. Hughes, ``{The Basics of gravitational wave theory},''
  {\em New J.Phys.} {\bf 7} (2005) 204,
  \href{http://xxx.lanl.gov/abs/gr-qc/0501041}{{\tt gr-qc/0501041}}.

\bibitem{Jaccard:2012ut}
M.~Jaccard, M.~Maggiore, and E.~Mitsou, ``{Bardeen variables and hidden gauge
  symmetries in linearized massive gravity},'' {\em Phys.Rev.} {\bf D87} (2013)
  044017, \href{http://xxx.lanl.gov/abs/1211.1562}{{\tt 1211.1562}}.

\end{thebibliography}\endgroup
